\documentclass[useAMS,usenatbib]{mn2e}

 \usepackage{Times}

\usepackage{psfrag}
\usepackage{pspicture}
\usepackage{graphicx}

\newcommand\lsim{~\lower.5ex\hbox{$\buildrel < \over \sim$}~}
\newcommand\gsim{~\lower.5ex\hbox{$\buildrel > \over \sim$}~}

\def\gsim{~\lower.6ex\hbox{$\buildrel > \over \sim$}~}

\title[11 Gyr evolution of H$\alpha$ SF Galaxies]{A large H$\alpha$ survey at $\bf z=2.23,1.47,0.84 \, \&\, 0.40$: the 11\,Gyr evolution of star-forming galaxies from HiZELS\thanks{This work is based on observations obtained using the Wide Field CAMera (WFCAM) on the 3.8m United Kingdom Infrared Telescope (UKIRT), as part of the High-redshift(Z) Emission Line Survey (HiZELS; U/CMP/3 and U/10B/07). It also relies on observations conducted with HAWK-I on the ESO Very Large Telescope (VLT), program 086.7878.A, and observations obtained with Suprime-Cam on the Subaru telescope (S10B-144S).} }
\author[D. Sobral et al.]{David Sobral$^{1}$\thanks{NOVA Fellow; E-mail: sobral@strw.leidenuniv.nl}, Ian Smail$^{2}$, Philip N. Best$^{3}$, James E. Geach$^{4}$, Yuichi Matsuda$^{5}$,
\newauthor John P. Stott$^{2}$, Michele Cirasuolo$^{4,6}$ \& Jaron Kurk$^{7}$ \\
$^{1}$ Leiden Observatory, Leiden University, P.O.\ Box 9513, NL-2300 RA Leiden, The Netherlands\\
$^{2}$ Institute for Computational Cosmology, Durham University, South Road, Durham, DH1 3LE, UK\\
$^{3}$ SUPA, Institute for Astronomy, Royal Observatory of Edinburgh, Blackford Hill, Edinburgh, EH9 3HJ, UK\\
$^{4}$ Department of Physics, McGill University, Ernest Rutherford Building, 3600 Rue University, Montr\'eal, Qu\'ebec, Canada, H3A 2T8\\
$^5$ Cahill Center for Astronomy \& Astrophysics, California Institute of Technology, MS 249-17, Pasadena, CA 91125, USA \\
$^6$ UK Astronomy Technology Centre, Royal Observatory, Blackford Hill, Edinburgh EH9 3HJ\\
$^7$ Max-Planck-Institut f{\"u}r Astrophysik, Karl-Schwarzschild Strasse 1, D-85741 Garching, Germany }
\begin{document}
\date{Accepted 2012 September 27.  Received 2012 September 27; in original form 2012 February 15}
\pagerange{\pageref{firstpage}--\pageref{lastpage}} \pubyear{2012}
\maketitle

\label{firstpage}
\begin{abstract}
This paper presents new deep and wide narrow-band surveys undertaken with UKIRT, Subaru and the VLT; a unique combined effort to select large, robust samples of H$\alpha$ star-forming galaxies at $z=0.40$, $0.84$, $1.47$ and $2.23$ (corresponding to look-back times of 4.2, 7.0, 9.2 and 10.6 Gyrs) in a uniform manner over $\sim2$ deg$^2$ in the COSMOS and UDS fields. The deep multi-epoch H$\alpha$ surveys reach a matched 3\,$\sigma$ flux limit  of $\approx3$\,M$_{\odot}$\,yr$^{-1}$ out to $z=2.2$ for the first time, while the wide area and the coverage over two independent fields allow to greatly overcome cosmic variance and assemble by far the largest samples of H$\alpha$ emitters. Catalogues are presented for a total of 1742, 637, 515 and 807 H$\alpha$ emitters, robustly selected at $z=0.40$, $0.84$, $1.47$ and $2.23$, respectively, and used to determine the H$\alpha$ luminosity function and its evolution. The faint-end slope of the H$\alpha$ luminosity function is found to be $\alpha=-1.60\pm0.08$ over $z=0-2.23$, showing no significant evolution. The characteristic luminosity of SF galaxies, $L_{\rm H\alpha}^*$, evolves significantly as $\log\,L^*_{\rm H\alpha}(z)=0.45z+\log\,L^*_{z=0}$. This is the first time H$\alpha$ has been used to trace SF activity with a single homogeneous survey at $z=0.4-2.23$. Overall, the evolution seen with H$\alpha$ is in good agreement with the evolution seen using inhomogeneous compilations of other tracers of star formation, such as FIR and UV, jointly pointing towards the bulk of the evolution in the last 11\,Gyrs being driven by a statistically similar SF population across cosmic time, but with a strong luminosity increase from $z\sim0$ to $z\sim2.2$. Our uniform analysis allows to derive the H$\alpha$ star formation history of the Universe (SFRH), showing a clear rise up to $z\sim2.2$, for which the simple parametrisation $\log_{10}\rho_{\rm SFR}=-2.1(1+z)^{-1}$ is valid over 80 per cent of the age of the Universe. The results reveal that both the shape and normalisation of the H$\alpha$ SFRH are consistent with the measurements of the stellar mass density growth, confirming that our H$\alpha$ SFRH is tracing the bulk of the formation of stars in the Universe for $z<2.23$. The star formation activity over the last $\sim$\,11\,Gyrs is responsible for producing $\sim95$ per cent of the total stellar mass density observed locally, with half of that being assembled in 2\,Gyrs between $z=1.2$--$2.2$, and the other half in 8\,Gyrs (since $z<1.2$). If the star-formation rate density continues to decline with time in the same way as seen in the past $\sim11$\,Gyrs, then the stellar mass density of the Universe will reach a maximum which is only 5 per cent higher than the present-day value.

\end{abstract}


\section{Introduction}\label{intro}

Observational studies show that star formation activity in galaxies, as measured through the star formation rate density ($\rho_{\rm SFR}$) in the Universe, has been decreasing significantly with time \citep[e.g.][]{Lilly96}.  Nevertheless, while surveys reveal that $\rho_{\rm SFR}$ rises steeply out to at least $z\sim1$ \citep[e.g.][]{Hopkins2006}, determining the redshift where $\rho_{\rm SFR}$ might have peaked at $z>1$ is still an open problem. This is because the use of different techniques/indicators (affected by different biases, dust extinctions and with different sensitivities -- and that can only be used over limited redshift windows) results in a very blurred and scattered understanding of the star formation history of the Universe. Other problems/limitations result from the difficulty of obtaining both large-area, large-sample, clean and deep observations (to overcome both cosmic variance, and avoid large extrapolations down to faint luminosities).

One way to make significant progress in our understanding of star formation at high redshifts is through the use of narrow-band imaging techniques. These can provide sensitive wide-field surveys to select star-forming galaxies through a single emission line and track it out to high redshift as it shifts from the optical into the near-IR. While there are a number of emission lines which are used to trace star formation, H$\alpha$ is by far the best at $z<3$,\footnote{It may be possible to extend H$\alpha$ studies to even higher redshifts: \cite{Shim} suggest that {\it Spitzer} IRAC mid-IR fluxes can be used to detect strong H$\alpha$ emission at even higher redshifts ($z\sim4$). NIRCAM and NIRISS on the {\it James Webb Space Telescope} will obviously significantly expand the exploration of H$\alpha$ emission at such high redshifts.} as it provides a sensitive census of star formation activity, is well-calibrated, and suffers only modest extinction in typical star-forming galaxies \citep[e.g.][]{Gilbank,Garn2010a,Sobral11B}, in contrast to shorter-wavelength emission lines. Furthermore, H$\alpha$ is also a much better estimate of the instantaneous star formation rate when compared to other widely used tracers, such as UV, FIR, or radio, as it sensitive to the presence of the most massive stars only, which are very short-lived. Even longer wavelength star-formation tracing emission lines, such as the Paschen series lines, are less affected by dust extinction, but they are intrinsically fainter than H$\alpha$ (e.g.\ Pa$\alpha$ is intrinsically $\sim10\times$ weaker than H$\alpha$ for a typical star-forming galaxy) and hence provide much less sensitive surveys out to lower redshifts.

H$\alpha$ surveys have been carried out by many authors \citep[e.g.][]{Bunker95,Malkan1996}, but they initially resulted in a relatively low number of sources for $z>0.5$ surveys. Fortunately, the development of wide field near-IR detectors has recently allowed a significant increase in success: at $z\sim2$, narrow-band surveys such as \cite{Moorwood}, which could only detect a handful of emitters, have been rapidly extended by others, such as \cite{G08}, increasing the sample size by more than an order of magnitude. Substantial advances have also been obtained at $z\sim1$ \citep[e.g.][]{Villar,S09a,CHU11}. Other H$\alpha$ surveys have used dispersion prisms on {\sc hst} to make progress \citep[e.g.][]{McCarthy,Yan,Hopkins2000,Shim09}, and there is promising work being conducted using the upgraded WFC3-grism \citep[e.g. WISP or 3D-HST;][]{Atek2010,HSTTHER,vanDokkum}.

%
%
\begin{table}
 \centering
  \caption{Narrow-band filters used to conduct the multi-epoch surveys for H$\alpha$ emitters, indicating the central wavelength ($\umu$m), full width at half maximum (FHWM), the redshift range for which the H$\alpha$ line is detected over the filter FWHM, and the corresponding volume (per square degree) surveyed (for the H$\alpha$ line). Note that the NB921 filter provides an [O{\sc ii}] survey which precisely matches the H$\alpha$ $z=1.47$ survey, and also a [O{\sc iii}] survey which broadly matches the $z=0.84$ H$\alpha$ survey. The NB$_{\rm J}$ and NB$_{\rm H}$ filters also provide [O{\sc ii}]\,3727 and [O{\sc iii}]\,5007 surveys, respectively, which match the $z=2.23$ NB$_{\rm K}$ H$\alpha$ survey.}
  \begin{tabular}{@{}ccccc@{}}
  \hline
  \bf NB filter  & $\bf \lambda_{\rm c}$ & \bf FWHM &$z$ H$\alpha$ & \bf Volume (H$\alpha$)  \\
   & ($\umu$m) & (\AA) & & ($10^4$\,Mpc$^3$\,deg$^{-2}$) \\
 \hline
   \noalign{\smallskip}
NB921 & 0.9196 &132 & 0.401$\pm$0.010 &  5.13 \\
NB$_{\rm J}$ & 1.211 & 150 & 0.845$\pm$0.015 & 14.65 \\
NB$_{\rm H}$ &1.617 & 211 & 1.466$\pm$0.016 & 33.96 \\
NB$_{\rm K}$ & 2.121 & 210  & 2.231$\pm$0.016 & 38.31 \\
HAWK-I H$_2$ & 2.125 & 300 &  2.237$\pm$0.023 & 54.70 \\
 \hline
\end{tabular}
\label{numbers}
\end{table}

HiZELS, the High-redshift(Z) Emission Line Survey\footnote{For more details on the survey, progress and data releases, see http://www.roe.ac.uk/ifa/HiZELS/} \citep[][hereafter S09 and S12]{G08,S09a,Sobral11B} is a Campaign Project using the Wide Field CAMera (WFCAM) on the United Kingdom Infra-Red Telescope (UKIRT), as well as the Suprime-Cam on the Subaru Telescope and the HAWK-I camera on VLT.  On UKIRT, HiZELS exploits specially-designed narrow-band filters in the $J$ and $H$ bands (NB$_{\rm J}$ and NB$_{\rm H}$), along with the H$_2$S(1) filter in the $K$ band (hereafter NB$_{\rm K}$), to undertake panoramic,  deep surveys for line emitters.   The Subaru observations provide a comparable survey at $z=0.40$ using the NB921 filter on Suprime-Cam, while the HAWK-I observations extend the UKIRT survey to fainter limits at $z=2.23$ over a smaller area. The combined elements of HiZELS primarily target the H$\alpha$ emission line \citep[but also other lines e.g.][]{S09b} redshifted into the red or near-infrared at $z=0.40$, $z=0.84$, $z=1.47$ and $z=2.23$ \citep[see][]{Best2010}, while the NB$_{\rm J}$ and NB$_{\rm H}$ filters also detect [O{\sc ii}]\,3727 and [O{\sc iii}]\,5007 emitters at $z=2.23$, matching the NB$_{\rm K}$ H$\alpha$ coverage at the same redshift.

%
\begin{table*}
 \centering
  \caption{Observation log of all the narrow-band observations obtained over the COSMOS and UDS fields, taken using WFCAM on UKIRT, HAWK-I on the VLT, and Suprime-Cam on Subaru, during 2006--2012. Limiting magnitudes (3\,$\sigma$, all in Vega) are the average of each field/pointing, based on the measurements of 10$^4$-10$^6$ randomly placed 2$''$ apertures in each frame, but note that for COSMOS (NB$_{\rm H}$ and NB$_{\rm K}$) the central 4 pointings overlap with part of the outer pointings (see Figure 2), making the central region data up to 0.3-0.4 mag deeper, by doubling/tripling the exposure time.}
  \begin{tabular}{@{}cccccccccc@{}}
  \hline
  \bf Field & \bf Band & \bf R.A. & \bf Dec. & \bf Int.\ time & \bf FHWM  & \bf Dates  &\bf $\bf m_{lim}$(Vega)  \\
        & (filter) & {(J2000)} &(J2000) & (ks) & ($''$) &  & (3$\sigma$) \\
 \hline
   \noalign{\smallskip}
COSMOS-1 & NB921 & 09\,59\,23 & $+$02\,30\,29 & 2.9 & 0.9 & 2010 Dec 9 & 24.4 \\
COSMOS-2 & NB921 & 10\,01\,34 & $+$02\,30\,29  &  2.9 & 0.9 & 2010 Dec 9 &24.5 \\
COSMOS-3 & NB921 & 09\,59\,23 & $+$02\,04\,16  &  2.9 & 0.9 & 2010 Dec 9 & 24.5  \\
COSMOS-4 &  NB921 & 10\,01\,34 & $+$02\,04\,16  &  2.9 & 0.9 & 2010 Dec 9 & 24.5  \\
UKIDSS-UDS C & NB921 & 02\,18\,00 & $-$05\,00\,00 & 30.0 & 0.8 & 2005 Oct 29, Nov 1, 2007 Oct 11$-$12 & 26.6  \\
UKIDSS-UDS N  & NB921 & 02\,18\,00 & $-$04\,35\,00 & 37.8 & 0.9 & 2005 Oct 30,31, Nov 1, 2006 Nov 18, 2007 Oct 11,12 & 26.7  \\
UKIDSS-UDS S & NB921 & 02\,18\,00 & $-$05\,25\,00 & 37.1 & 0.8 & 2005 Aug 29, Oct 29, 2006 Nov 18, 2007 Oct 12 & 26.6  \\
UKIDSS-UDS E & NB921 & 02\,19\,47 & $-$05\,00\,00 & 29.3 & 0.8 & 2005 Oct 31, Nov 1, 2006 Nov 18, 2007 Oct 11,12 & 26.6  \\
UKIDSS-UDS W & NB921 & 02\,16\,13 & $-$05\,00\,00 & 28.1 & 0.8 & 2006 Nov 18, 2007 Oct 11,12 & 26.0  \\
 \hline
 COSMOS-NW(1) & NB$_{\rm J}$ & 10\,00\,00 & +02\,10\,30 & 19.7 & 0.8 & 2007 Jan 14--16  & 22.0  \\
 COSMOS-NE(2) &  NB$_{\rm J}$ & 10\,00\,52 & +02\,10\,30 & 23.8 & 0.9 & 2006 Nov 10; 2007 Jan 13--14 & 22.0 \\
 COSMOS-SW(3) &  NB$_{\rm J}$ & 10\,00\,00 & +02\,23\,44 & 18.9 & 0.9 & 2007 Jan 15--17 & 22.0 \\
 COSMOS-SE(4) &  NB$_{\rm J}$ & 10\,00\,53 & +02\,23\,44 & 17.1 & 1.0 & 2007 Jan 15, 17; Feb  13, 14, 16 & 21.9 \\
 UKIDSS-UDS  NE & NB$_{\rm J}$ & 02\,18\,29 & $-$04\,52\,20 & 20.9 & 0.8 & 2007 Oct  21--23 & 22.0 \\
UKIDSS-UDS  NW & NB$_{\rm J}$ & 02\,17\,36 & $-$04\,52\,20 & 22.4 & 0.9 & 2007 Oct 20--21 & 22.1 \\
UKIDSS-UDS  SE & NB$_{\rm J}$ & 02\,18\,29 & $-$05\,05\,53 & 19.6 & 0.9 & 2007 Oct  23, 24 & 22.0  \\
UKIDSS-UDS  SW & NB$_{\rm J}$ &  02\,17\,38 & $-$05\,05\,34 & 22.4 & 0.8 & 2007 Oct 19, 21 & 22.0  \\
 \hline
COSMOS-NW(1) & NB$_{\rm H}$ & 10\,00\,00 & $+$02\,10\,30 & 12.5 & 1.0 & 2009 Feb 27; Mar 1-2 & 21.1 \\
COSMOS-NE(2) DEEP & NB$_{\rm H}$ & 10\,00\,52 & $+$02\,10\,30 & 107.0 & 0.9 & 2009 Feb 28; Apr 19; May 22; 2011 Jan 26--30 & 22.2  \\
COSMOS-SW(3) & NB$_{\rm H}$ & 10\,00\,00 & $+$02\,23\,44 & 14.0 & 0.7 & 2010 Apr 2 & 21.0 \\ 
COSMOS-SE(4) & NB$_{\rm H}$ & 10\,00\,53 & $+$02\,23\,44 & 18.1 & 1.0 & 2009 Mar 2; Apr 30; May 22; 2010 Apr 3 & 20.8 \\ 
COSMOS-A & NB$_{\rm H}$ & 10\,00\,01 & $+$02\,36\,53 & 12.6 & 1.0 & 2010 Apr 9; 2011 Jan 25 & 21.1 \\
COSMOS-B & NB$_{\rm H}$ & 10\,00\,54 & $+$02\,36\,30 &12.6 & 0.9 & 2010 Apr 8-9 & 21.0 \\
COSMOS-C & NB$_{\rm H}$ & 10\,00\,01 & $+$01\,57\,10 & 13.0 & 0.8 & 2010 Apr 6-8 & 20.8  \\
COSMOS-D & NB$_{\rm H}$ & 10\,00\,52 & $+$01\,57\,15  & 14.2 & 0.8 & 2010 Apr 7-8 & 21.0  \\
COSMOS-E & NB$_{\rm H}$& 09\,59\,07 & $+$02\,23\,44  & 12.6 & 0.8 & 2010 Apr 3-6 & 20.9 \\
COSMOS-F & NB$_{\rm H}$ & 09\,59\,07 & $+$02\,10\,30  & 12.6 & 0.7 & 2010 Apr 4 & 21.1\\
COSMOS-G & NB$_{\rm H}$ & 10\,01\,46 & $+$02\,23\,44  & 14.5 & 0.8 & 2010 Apr 4 \& 9 & 21.0  \\
COSMOS-H  &  NB$_{\rm H}$ & 10\,01\,48 & $+$02\,10\,51  & 14.0 & 0.8 & 2010 Apr 6 \& 9 & 20.8  \\
UKIDSS-UDS NE & NB$_{\rm H}$ & 02\,18\,29 & $-$04\,52\,20 & 18.2 & 0.9 & 2008 Sep 28-29; 2009 Aug 16-17; 2010 Jul 22 & 21.3 \\
UKIDSS-UDS NW & NB$_{\rm H}$ & 02\,17\,36 & $-$04\,52\,20 & 18.0 & 0.9 & 2008 Sep 25, 29; 2010 Jul 18, 22 & 20.9 \\
UKIDSS-UDS SE & NB$_{\rm H}$ & 02\,18\,29 & $-$05\,05\,53 & 25.2 & 0.8 & 2008 Sep 25, 28-29; 2009 Aug 16-17 & 21.5  \\
UKIDSS-UDS SW  & NB$_{\rm H}$ &  02\,17\,38 & $-$05\,05\,34 & 19.6 & 0.9 & 2008 Oct-Nov; 2009 Aug 16-17; 2010 Jul 23 & 21.3  \\
 \hline
COSMOS-NW(1) & NB$_{\rm K}$ & 10\,00\,00 & $+$02\,10\,30 & 24.3 & 0.9 & 2006 Dec 17, 19--20; 2008 May 11; 2009 Feb 27 & 21.0  \\
COSMOS-NE(2) DEEP & NB$_{\rm K}$ & 10\,00\,52 & $+$02\,10\,30 & 62.5 & 0.9 & 2006 May 20--21; Dec 20; 2008 Mar 6-9 & 21.3 \\
COSMOS-SW(3) & NB$_{\rm K}$ & 10\,00\,00 & $+$02\,23\,44 & 20.0 & 0.9 & 2006 May 22, 24, Dec 20; 2009 May 20 & 20.8 \\
COSMOS-SE(4) & NB$_{\rm K}$ & 10\,00\,53 & $+$02\,23\,44 & 19.5 & 0.9 & 2006 Nov 13--15, 30; Dec 16 & 20.9 \\
COSMOS-A &  NB$_{\rm K}$ & 10\,00\,01 & $+$02\,36\,53 & 20.0 & 1.0 & 2011 Mar 19-26; 2012 Mar 20 & 21.0  \\
COSMOS-B &  NB$_{\rm K}$ & 10\,00\,54 & $+$02\,36\,30 & 20.0 & 0.9 & 2011 Mar 25-26 & 20.8 \\
COSMOS-C & NB$_{\rm K}$ & 10\,00\,01 & $+$01\,57\,10 & 26.7 & 0.9 & 2011 Mar 27, 30; Apr 3-5, 16-18 & 20.9  \\
COSMOS-D & NB$_{\rm K}$ & 10\,00\,52 & $+$01\,57\,15  & 20.0 & 0.9 & 2011 Mar 30; Jun 6; 2012 Jan 5,18; Feb 26; Mar 2 & 20.9  \\
COSMOS-E &  NB$_{\rm K}$ & 09\,59\,07 & $+$02\,23\,44  & 20.0 & 0.8 & 2011 Mar 30; May 18-23, 30 & 20.7 \\
COSMOS-F & NB$_{\rm K}$ & 09\,59\,07 & $+$02\,10\,30  & 20.0 & 0.9 & 2011 Dec 18; 2012 Mar 2, 17 & 20.8\\
COSMOS-G & NB$_{\rm K}$ & 10\,01\,46 & $+$02\,23\,44  & 20.0 & 0.9 & 2012 Mar 18-19 & 20.9  \\
COSMOS-H  &  NB$_{\rm K}$ & 10\,01\,48 & $+$02\,10\,51  & 20.0 & 0.8 & 2012 Mar 19-20  & 20.7  \\
UKIDSS-UDS NE & NB$_{\rm K}$ & 02\,18\,29 & $-$04\,52\,20 & 19.2 & 0.8 & 2005 Oct 18; 2006 Nov 13-14  & 20.7 \\
UKIDSS-UDS NW & NB$_{\rm K}$ &  02\,17\,36 & $-$04\,52\,20 & 20.0 & 0.9 & 2006 Nov 11 & 20.8 \\
UKIDSS-UDS SE & NB$_{\rm K}$ &  02\,18\,29 & $-$05\,05\,53 & 18.7 & 0.8 & 2006 Nov 15-16 & 20.7  \\
UKIDSS-UDS SW  &  NB$_{\rm K}$ &  02\,17\,38 & $-$05\,05\,34 & 23.6 & 0.8 & 2007 Sep 30 & 20.9 \\
 \hline
COSMOS-HAWK-I & H2 & 10\,00\,00 & $+$02\,10\,30  & 19.4 & 0.9 & 2009 Apr 10, 14-15, 18; May 13-14 & 21.5 \\
UKIDSS-UDS-HAWK-I & H2 & 02\,17\,36 & $-$04\,52\,20 & 19.1 & 1.0 & 2009 Aug 16, 19, 24, 27 & 21.7 \\

 \hline
\end{tabular}
\label{obs}
\end{table*}

One of the main aims of HiZELS is to provide measurements of the evolution of the H$\alpha$ luminosity function from $z=0.0$ to $z=2.23$ \citep[but also other properties, such as clustering, environment and mass dependences; c.f.][]{SOBRAL10A,SOBRAL10B,GEACH12}. The first results \citep[][S09; S12]{G08} indicate that the H$\alpha$ luminosity function evolves significantly, mostly due to an increase of about one order of magnitude in L$_{\rm H\alpha}^*$, the characteristic H$\alpha$ luminosity, from the local Universe to $z=2.23$ (S09). In addition, \cite{SOBRAL10B} found that at $z=0.84$ the faint-end slope of the luminosity function ($\alpha$) is strongly dependent on the environment, with the H$\alpha$ luminosity function being much steeper in low density regions and much shallower in the group/cluster environments.

However, even though the progress has been quite remarkable, significant issues remain to be robustly addressed for a variety of reasons. For example, is the faint-end slope of the H$\alpha$ luminosity function ($\alpha$) becoming steeper from low to high redshift? Results from \cite{Hayes} point towards a steep faint-end slope at $z>2$. However, Hayes et al. did not sample the bright end, and have only targeted one single field over a relatively small area, and thus cosmic variance could play a huge role. \cite{Tadaki} find a much shallower $\alpha$ at $z\sim2$ using Subaru. Furthermore, measurements so far rely on different data, obtained down to different depths and using different selection criteria. Additionally, different ways of correcting for completeness \citep[c.f. for example][]{CHU11}, filter profiles or contamination by the [N{\sc ii}]$_{\lambda\lambda6548,6583.6}$ lines can also lead to significant differences. How much of the evolution is in fact real, and how much is a result of different ways of estimating the H$\alpha$ luminosity function? This can only be fully quantified with a completely self-consistent multi-epoch selection and analysis. Another issue which still hampers the progress is overcoming cosmic variance and probing a very wide range of environments and stellar masses at $z>1$. Large samples of homogeneously selected star-forming galaxies at different epochs up to $z>2$ would certainly be ideal to provide strong tests on our understanding of how galaxies form and how they evolve.

%
%
\begin{figure}
\centering
\includegraphics[width=8.2cm]{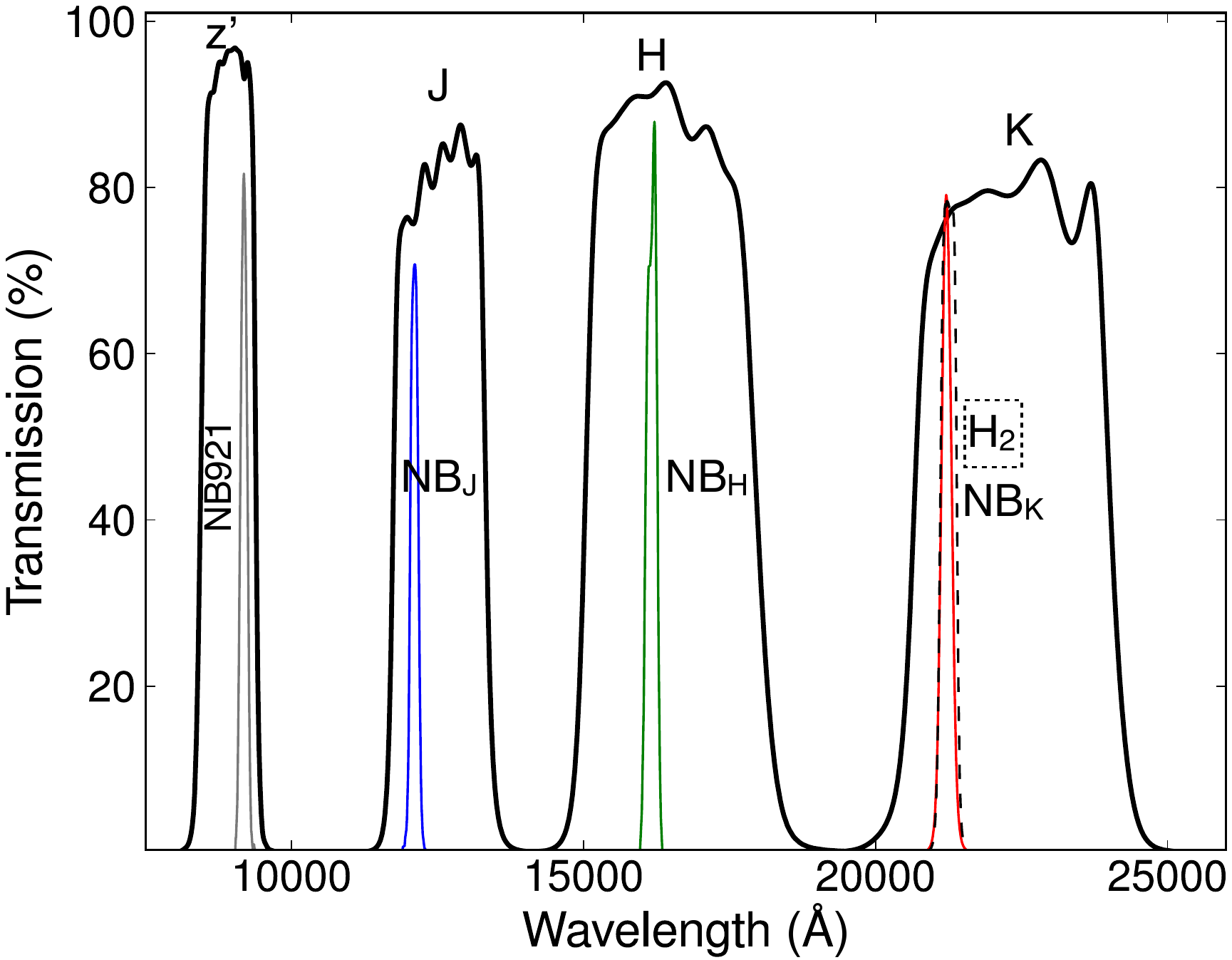}
\caption[FILTERS]{The broad- and narrow-band filter profiles used for the analysis. The narrow-band filters in the $z'$, $J$, $H$ and $K$ bands (typical FWHM of $\approx100-200$\,\AA) trace the redshifted H$\alpha$ line at $z=0.4,0.84,1.47,2.23$ very effectively, while the (scaled) broad-band imaging is used to estimate and remove the contribution from the continuum. Note that because the filters are not necessary located at the center of the respective broad-band transmission profile, very red/blue sources can produce narrow-band excesses which mimic emission lines; that is corrected by estimating the continuum colour of each source and correcting for it. \label{line_frac_fluxes}}
\end{figure}

In order to clearly address the current shortcomings and provide the data that is required, we have undertaken by far the largest area, deep multi-epoch narrow-band H$\alpha$ surveys over two different fields. By doing so, both faint and bright populations of equally selected H$\alpha$ emitters at $z=0.4$, $0.84$, $1.47$ and $2.23$ have been obtained, using 4 narrow band filters (see Figure 1 and Table 1). This paper presents the narrow-band imaging and results obtained with the NB921, NB$_{\rm J}$, NB$_{\rm H}$, NB$_{\rm K}$ and H$_2$ filters on Subaru, the United Kingdom InfraRed Telescope (UKIRT) and the Very Large Telescope (VLT), over a total of $\sim2\deg^2$ in the Cosmological Evolution Survey \citep[COSMOS;][]{Scoville} and the SXDF Subaru-XMM--UKIDSS Ultra Deep Survey \citep[UDS;][]{Lawrence} fields.

The paper is organised as follows: \S2 describes the observations, data reduction, source extraction, catalogue production, selection of line emitters, and the samples of H$\alpha$ emitters. In \S3, after estimating and applying the necessary corrections, the H$\alpha$ luminosity functions are derived at $z=0.4$, $0.84$, $1.47$ and $2.23$, together with an accurate measurement of their evolution at both bright and faint ends. In \S4 the star formation rate density at each epoch is also evaluated and the star formation history of the Universe is presented. \S4 also discusses the results in the context of galaxy formation and evolution in the last 11\,Gyrs, including the inferred stellar mass density growth. Finally, \S5 presents the conclusions. An H$_0=70$\,km\,s$^{-1}$\,Mpc$^{-1}$, $\Omega_M=0.3$ and $\Omega_{\Lambda}=0.7$ cosmology is used. Narrow-band magnitudes in the near-infrared and the associated broad-band magnitudes are in the Vega system, except when noted otherwise (e.g. for colour-colour selections). NB921 and $z'$ magnitudes are given in the AB system (except in Table 2, where they are given in Vega for direct comparison).

\section{DATA AND SAMPLES}\label{data_technique}

\subsection{Optical NB921 imaging with Subaru}\label{observationes_NB921}

Optical imaging data were obtained with Suprime-Cam using the NB921 narrow-band filter. Suprime-Cam consists of 10 CCDs with a combined field of view of $34'\times27'$ and with chip gaps of $\sim15''$. The NB921 filter is centered at 9196\,\AA \ with a FWHM of 132\,\AA. The COSMOS field was observed in service mode in December 2010 with four different pointings covering the central 1.1\,deg$^2$. Total exposure times were 2.9\,ks per pointing, composed of individual exposures of 360\,s dithered over 8 different positions. Observations are detailed in Table 2. The UDS field has also been observed with the NB921 filter \citep[see][]{Ouchi10}, and these data have been extracted from the archive. Full details of the data reduction and catalogue production of the UDS data were presented by S12 and the same approach was adopted for the COSMOS data. In brief, all the raw NB921 data were reduced with the Suprime-Cam Deep field REDuction package \citep[{\sc sdfred},][]{Yagi2002,Ouchi2004} and {\sc iraf}. The combined images were aligned to the public $z'$-band images of Subaru-XMM Deep Survey or the COSMOS field and PSF matched (FWHM$=0.9''$). The NB921 zero points were determined using $z'$ data, so that the ($z'$-NB921) colours are consistent with a median of zero for $z'$ between 19 and 21.5 -- where both NB921 and $z'$ images are unsaturated and have very high signal-to-noise ratios.

Source detection and photometry were performed using {\sc SExtractor} \citep{SExtractor}. Sources were detected on each individual NB921 image and magnitudes measured with $2''$ and $3''$ diameter apertures. The $3''$ apertures are used to select and measure H$\alpha$ line fluxes: at $z=0.4$ the 3$''$ apertures measure the same physical area as 2$''$ apertures at $z=0.8,1.47,2.23$ ($\approx16$\,kpc), assuring full consistency. The $2''$ apertures are used to measure emission lines from sources at higher redshift ([O{\sc ii}] at $z=1.47$, to match the NB$_{\rm H}$ H$\alpha$ measurement at the same redshift, and [O{\sc iii}] at $z=0.84$ to match the NB$_{\rm J}$ H$\alpha$ survey). The average NB921 3$\sigma$ limiting magnitudes (in 2$''$ apertures) are given in Table \ref{obs}. 

%
%
\begin{figure}
\centering
\includegraphics[width=8.2cm]{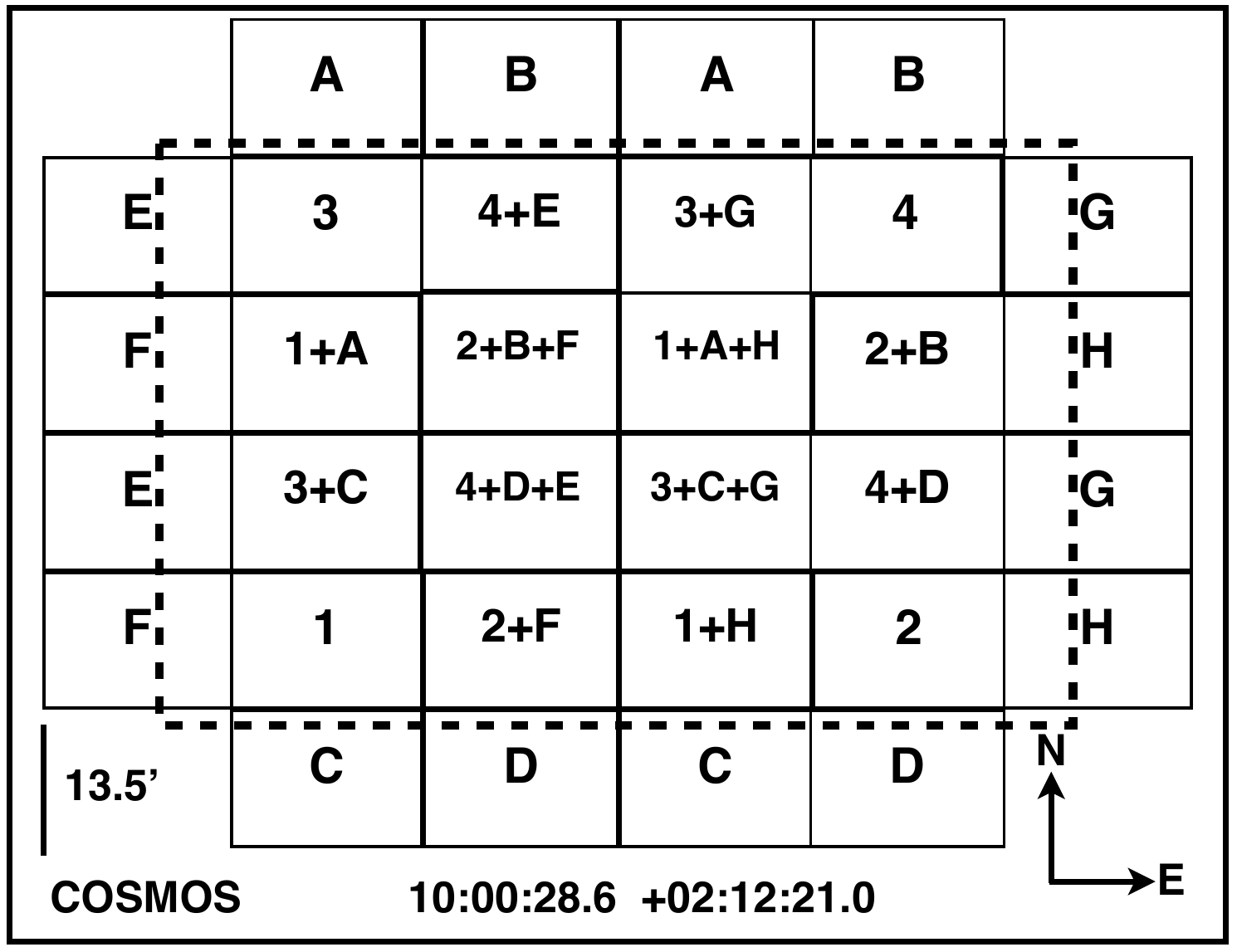}
\caption[FILTERS]{The survey strategy used to cover the COSMOS field. The central pointings (1,2,3,4; see Table 2) were complemented with further pointings (A to H) to both increase the surveyed area, and increase the exposure time in the central area. The region delimited by the dashed line shows the NB921 coverage obtained in COSMOS. See Table 2 for details on the pointings, including exposure times. \label{COSMOS_SURVEY}}
\end{figure}

\subsection{Near-infrared imaging with UKIRT}\label{observationes_UKIRT}

The COSMOS and UKIDSS UDS fields were observed with WFCAM on UKIRT as summarised in Table 2, using the NB$_{\rm J}$, NB$_{\rm H}$ and NB$_{\rm K}$  narrow-band filters, with central wavelengths and FWHM given in Table 1. WFCAM's has four $2048\times2048$ $0.4''$\,pixel$^{-1}$ detectors offset by $\sim20'$, resulting in a non-contiguous field of view of $\sim27'\times27'$ which can be macrostepped four times to cover a contiguous region of $\sim55'\times55'$. Observations were conducted over 2006--2012, covering $1.6$\,deg$^2$ (NB$_{\rm J}$) and $2.34$\,deg$^2$ (NB$_{\rm H}$ and NB$_{\rm K}$) over the COSMOS and the UDS fields (see Table 2).

The coverage over UDS is a simple mosaic obtained with 4 different pointings, covering a contiguous region of $\sim55'\times55'$. For the COSMOS field, an initial 0.8\,deg$^2$ coverage obtained with 4 WFCAM pointings was complemented in NB$_{\rm H}$ and NB$_{\rm K}$ by 8 further WFCAM pointings, macro-jittered to obtain a combined 1.6 \,deg$^2$ coverage with increasing exposure time per pixel towards the centre of the field (see Figure \ref{COSMOS_SURVEY}). Part of the central region ($\sim0.2$\,deg$^2$) benefits further from some significant extra deep data both in NB$_{\rm H}$ and NB$_{\rm K}$ (see Table 2), leading to a much higher total exposure time.

A dedicated pipeline has been developed for HiZELS (PfHiZELS, c.f. S09 for more details). The pipeline has been modified and updated since S09 mostly to 1) improve the flat-fielding\footnote{The improvements in the flat-fielding are obtained by stacking all second-pass flattened frames per field and producing source masks on the stacked images. Masks are then used to produce third pass flats using all the frames in the jitter sequence except the frame being flattened. The third-pass flattened frames are then stacked again, and the procedure is repeated another time. This procedure is able to both mask many sources which are undetected in individual frames out of the flats, but particularly to mask bright sources much more effectively, as the stacking of all images reveals a wider distribution of flux from those sources.} and 2) provide more accurate astrometric solutions for each individual frame which result in a more accurate stacking\footnote{{\sc iraf} and {\sc scamp} (Bertin et al. 2000) is used to distort correct the frames and obtain a very accurate ($rms\approx0.1-0.2''$) astrometric solution for each frame (using 2MASS), always assuring that the flux is conserved.}. The updated version of the pipeline (PfHiZELS2012) has been used to reduce all UKIRT narrow band data (NB$_{\rm J}$, NB$_{\rm H}$ and NB$_{\rm K}$), including those already presented in previous papers. This approach guarantees a complete self-consistency and takes advantage of the improved reduction which, in some cases, is able to go deeper by $\approx0.2$\,mag when compared to the data reduced by the previous version of the pipeline (e.g. S09).

For the COSMOS field, in order to co-add frames taken with different WFCAM cameras (due to the survey strategy, see Figure \ref{COSMOS_SURVEY}), {\sc scamp} is used (in combination with SDSS-DR7) to obtain accurate astrometry solutions which account for distortions in each stack (in addition to individual frames being corrected prior to combining) before co-adding different fields. The typical rms is $<0.1''$, by using on average $\sim500$ sources per chip. By following this approach, even at the highest radial distances ($r>1000$ pix) from the centre of the images the PSF/ellipticity remains unchanged due to stacking, and the data over areas that double/triple the expose time are found to become deeper by (on average) 0.3-0.4 mags, with no radial change in the PSF.

Narrow-band images were photometrically calibrated (independently) by matching $\sim100$ stars per frame with $J$, $H$ and $K$ between the 12th an 16th magnitudes from the 2MASS All-Sky catalogue of Point Sources \citep{2MASS} which are unsaturated in the narrow-band images. WFCAM images are affected by cross-talk and other artifacts caused by bright stars: accurate masks are produced in order to reject such regions. Sources were extracted using {\sc SExtractor} \citep{SExtractor}, making use of the masks. Photometry was measured in apertures of $2''$ diameter which at $z=0.8-2.2$ recover H$\alpha$ fluxes over $\approx16$\,kpc. The average 3\,$\sigma$ depths of the entire set of NB frames vary significantly, and are summarised in Table 2. The total numbers of sources detected with each filter are given in Table 3. Note that the central region of COSMOS NB$_{\rm H}$ and NB$_{\rm K}$ coverages benefits from a much higher total exposure time per pixel, resulting in data that are deeper by 0.3--0.4\,mag (on average) than the outer regions.

\subsection{Near-infrared H$_2$ imaging with HAWK-I}\label{observationes_HAWKI}

The UKIDSS UDS and COSMOS fields were observed with the HAWK-I instrument \citep{Pirard,Casali06} on the VLT during 2009. A single dithered pointing was obtained in each of the fields using the H$_2$ filter, characterised by $\lambda_c = 2.124\,\umu$m and $\delta\lambda = 0.030\,\umu$m (note that the filter is slightly wider than that on WFCAM). Individual exposures were of 60\,s, and the total exposure time per field is 5 hours. Table 2 presents the details of the observations and depth reached.

Data were reduced using the HAWK-I ESO pipeline recipes, by following an identical reduction scheme/procedure to the WFCAM data. The data have also been distortion corrected and astrometrically calibrated before combining, using the appropriate pipeline recipes. After combining all the individual reduced frames it is possible to obtain a contiguous image of $\approx7.5\times7.5$\,arc min$^2$ in each of the fields. There are, nonetheless, small regions with slightly lower exposure time per pixel in regions related with chip gaps at certain positions. Because of the availability of the very wide WFCAM imaging, regions in the HAWK-I combined images for which the exposure time per pixel is $<80$\% of the total are not considered. Frames are photometrically calibrated using 2MASS as a first pass, and then using UDS and COSMOS $K_s$ calibrated images to guarantee a median 0 colour ($K$-NB$_K$) for all magnitudes probed, as this procedure provides a larger number of sources. Similarly to the procedure used for WFCAM data, sources were extracted using {\sc SExtractor} and photometry was measured in apertures of $2''$ diameter. 

\subsection{Narrowband excess selection}\label{narrowB_exc_selection}

In order to select potential line emitters, broad-band ($BB$) imaging is used in the $z'$, $J$, $H$ and $K_s$ bands to match narrow-band ($NB$) imaging in the NB921, NB$_{\rm J}$, NB$_{\rm H}$ and NB$_{\rm K}$/H$_2$, respectively. Count levels on the broad-band images are scaled down to match the counts (of 2MASS sources) for each respective narrow-band image, in order to guarantee a median zero colour, and a common counts-to-magnitude zero point. Sources are extracted from $BB$ images using the same aperture sizes used for $NB$ images and matched to the $NB$ catalogue with a search radius of $<0.9''$. Note, however, that none of the narrow-band filters fall at the centre of the broad-band filters (see Figure 1). Thus, objects with significant continuum colours will not have $BB-NB=0$; this can be corrected with broad-band colours (c.f. S12), in order to guarantee that $BB-NB$ distribution is centred on 0 and has no dependence on continuum broad-band colours. Average colour corrections\footnote{Sources for which one of the broad-band colours is not available (typically 5 per cent of the sources) are assigned the median correction. The median corrections are: $-0.04$, $+0.07$, $+0.05$ and $+0.03$ for NB921, NB$_{\rm J}$, NB$_{\rm H}$ and NB$_{\rm K}$ respectively.} are given by:

\smallskip

\noindent $\rm(z'-NB921)_{AB}=(z'-NB921)_{0,AB}-0.05(J_{AB}-z'_{AB})-0.04$

\smallskip

\noindent $\rm(J-NB_{\rm J})=(J-NB_{\rm J})_0-0.09(z'_{AB}-J_{AB})+0.11$

\smallskip

\noindent $\rm(H-NB_{\rm H})=(H-NB_{\rm H})_0+0.07(J_{AB}-H_{AB})+0.06$

\smallskip

\noindent $\rm(K-NB_{\rm K})=(K-NB_{\rm K})_0-0.02(H_{AB}-K_{AB})+0.04$

\medskip

Potential line emitters are then selected according to the significance of their $(BB-NB)$ colour, as they will have $(BB-\,NB)>0$. True emitters are distinguished from those with positive colours due to the scatter in the magnitude measurements by quantifying the significance of the narrowband excess. The parameter $\Sigma$ \citep[see][]{Bunker95} quantifies the excess compared to the random scatter expected for a source with zero colour, as a function of narrow-band magnitude (see e.g. S09), and is given by:

\begin{equation}
\Sigma=\frac{1-10^{-0.4(BB-NB)}}{10^{-0.4(ZP-NB)}\sqrt{\pi r_{\rm ap}^2(\sigma_{\rm NB}^2+\sigma_{\rm BB}^2)}},
\end{equation}
where ZP is the zero point of the $NB$ (and $BB$, as those have been scaled to have the same ZP as $NB$ images), $r_{\rm ap}$ is the aperture radius (in pixels) used, and $\sigma_{\rm NB}$ and $\sigma_{\rm BB}$ are the rms (per pixel) of the $NB$ and $BB$ images, respectively.

Here, potential line emitters are selected if $\Sigma>3.0$ (see Figure 3). The spread on the brighter end (narrow-band magnitudes which are not saturated, but for which the scatter is not affected by errors in the magnitude, i.e., much brighter than the limit of the images) is quantified for each data-set and frame, and the minimum $(BB-NB)$ colour limit over bright magnitudes is set by the $3\times$ the standard deviation of the excess colour over such magnitudes ($s_b$). A common rest-frame EW limit of EW$_0=$\,25\,\AA \ is applied, guaranteeing a limit higher than the $3\times s_b$ dispersion over bright magnitudes in all bands. The combined selection criteria guarantees a clean selection of line emitters and, most importantly, it ensures that the samples of H$\alpha$ emitters are selected down to the same rest-frame EW, allowing one to quantify the evolution across cosmic time. An example of this selection for the full COSMOS NB$_{\rm K}$ data is shown in Figure 3 and the reader is referred to e.g. S09 and S12 for further examples.

%
%
\begin{figure}
\centering
\includegraphics[width=8.3cm]{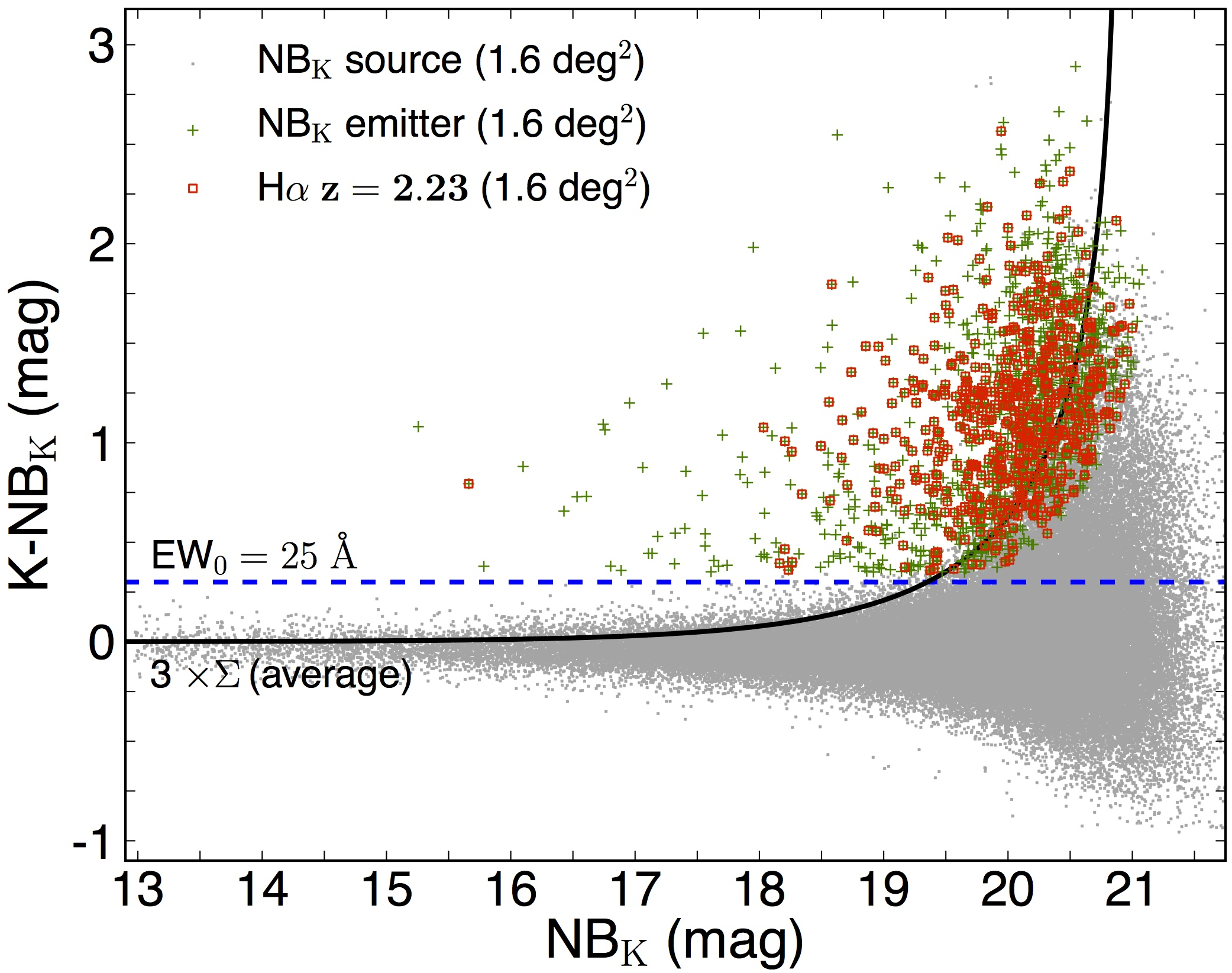}
\caption[The NBK excess]{Narrow-band excess as a function of narrow-band magnitude for NB$_{\rm K}$ (Vega magnitudes) data over the full COSMOS coverage (1.6\,deg$^2$). These show $>3\sigma$ detections in narrow-band imaging and the solid line present the average 3.0\,$\Sigma$ colour significance for NB$_{\rm K}$ (for the average depth, but note that the analysis is done individually for each frame, and that there are variations in depth ip to 0.6 mag difference). H$\alpha$ sources shown are narrow-band emitters which satisfy all the H$\alpha$ selection criteria fully described in Section 2.6. The horizontal dashed line presents the equivalent width cut used for NB$_{\rm K}$ data -- corresponding to a $z=2.23$ rest-frame EW limit of 25\,\AA \ for H$\alpha$+[N{\sc ii}]; this is the same rest-frame equivalent width used at all redshifts. \label{photoz}}
\end{figure}

As a further check on the selection criteria, the original imaging data are used to produce $BB$ and $NB$ postage stamp images of all the sources. The $BB$ is subtracted from the $NB$ image leaving the residual flux. From visual inspection these residual images contain obvious narrow band sources and it is found that the remaining flux correlates well with the catalogue significance.

\subsection{The samples of NB line emitters}\label{NB_emitters}

Narrow-band detections below the estimated 3\,$\sigma$ detection threshold were not considered. By using colour-colour diagnostics (see S12), potential stars are identified in the sample and rejected as well (the small fraction varies from band to band; see Table 3). The sample of remaining potential emitters ($\Sigma>3$ \& EW$_0>25$\,\AA; see Table 3 for numbers) is visually checked to identify spurious sources, artifacts which might not have been masked, or sources being identified in very noisy regions (see Table 3). Sources classed as spurious/artifacts are removed from the sample of potential emitters.  The final samples of line emitters are then derived. 

As a further test of the reliability of the line emitter samples, it can be noted that since the HAWK-I observations are both deeper and obtained over a larger redshift slice (due to a wider filter profile) when compared to WFCAM, they should be able to confirm all NB$_{\rm K}$ emitters over the matched area. This is confirmed, as all 10 emitters which are detected with WFCAM in the matched area are recovered by HAWK-I data as well.

The catalogues, containing all narrow-band emitter candidates, are presented in Appendix A. The catalogues provide IDs, coordinates, narrow-band and broad-band magnitudes, estimated fluxes and observed EWs. Further details and information are available on the HiZELS website.

%
%
\begin{figure*}
\centering
\includegraphics[width=16.8cm]{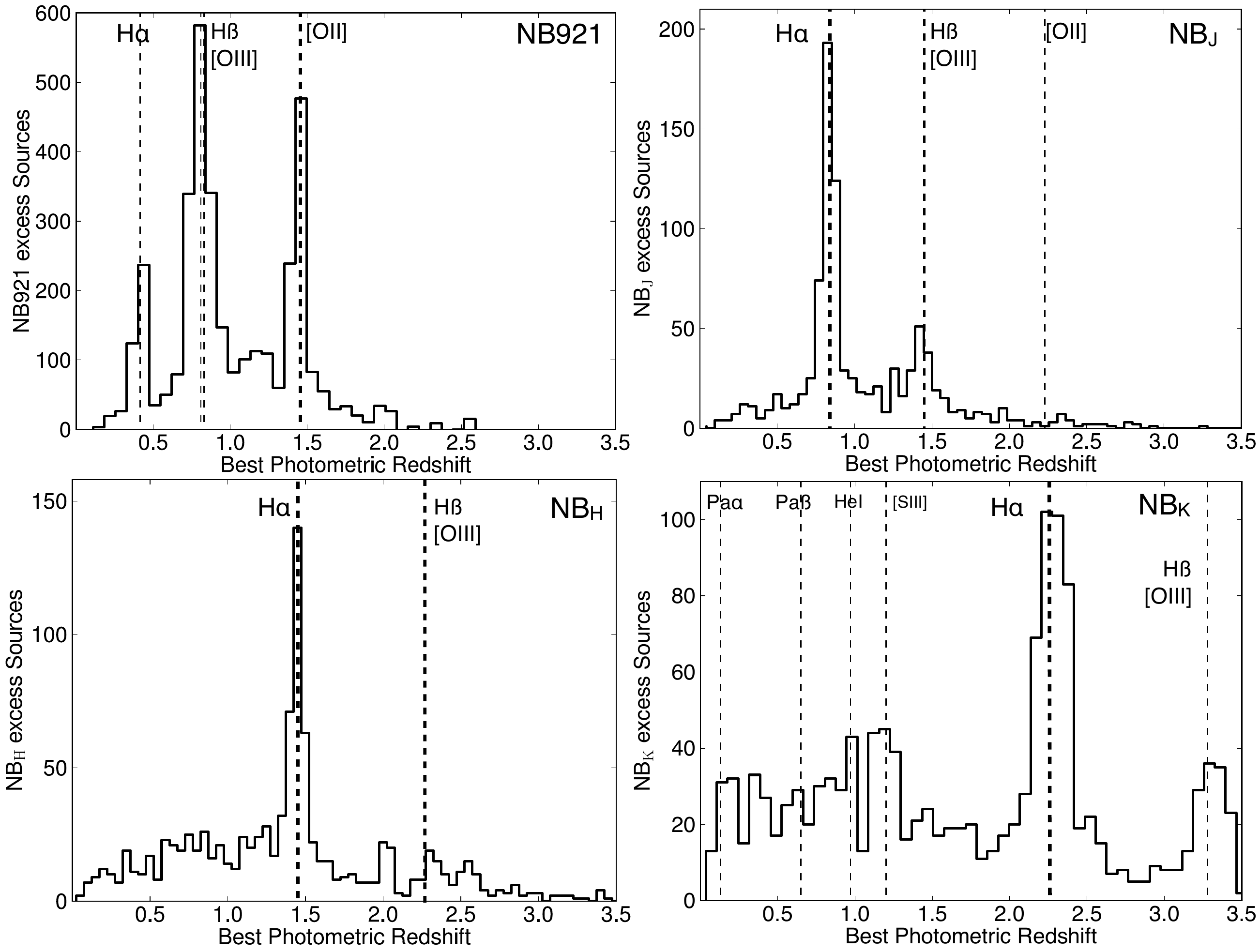}
\caption[The photometric redshifts distribution for the 3 narrow-band filters]{Photometric redshift distributions (peak of the probability distribution function for each source) for the NB921 line emitter candidates (upper left). NB$_{\rm J}$ (upper right), NB$_{\rm H}$ (lower left) and NB$_{\rm K}$ (lower right). All distributions peak at redshifts which correspond to strong emission lines, corresponding to H$\alpha$, H$\beta$, [O{\sc iii}] and [O{\sc ii}] emitters at various redshifts depending on the central wavelength of each narrow-band filter. Other populations of emitters are also found, such as Paschen-lines. The dotted lines indicate the redshift for which an emission line is detectable by the narrow-band filters. Ly$\alpha$ emitter candidates in the NB921 dataset (e.g. 18 sources in COSMOS) with photo-z of $z\sim6-7$ are not shown. \label{photoz}}
\end{figure*}

The photometric redshift \citep[Photo-$z$s from][]{Ilbert09,Cirasuolo10} distributions of the sources selected with the 4 narrow-band filters are presented in Figure \ref{photoz}. The photometric redshifts show clear peaks associated with H$\alpha$, H$\beta$/[O{\sc iii}]$_{\lambda\lambda 4959,5007}$, and [O{\sc ii}]$_{\lambda3727}$ (see Figure \ref{photoz}), together with further emission lines such as Paschen lines and Ly$\alpha$. Spectroscopic redshift are also available for a fraction of the selected line-emitters \citep[][]{Lilly09,Yamada05, Bart_Simpson06, Geach008,van_breu07, Ouchi2008, Smail08, Ono09}\footnote{See UKIDSS UDS website (http://www.nottingham.ac.uk/astronomy/UDS) for a redshift compilation by O. Almaini and the COSMOS data archive (http://irsa.ipac.caltech.edu/data/COSMOS) for the catalogues, spectra and information on the various instruments and spectroscopic programmes.} -- these will be discussed in the following Sections.

%
%
\begin{table*}
 \centering
  \caption{A summary of the number of sources, narrow-band emitters and H$\alpha$-selected emitters for the various surveys undertaken with different narrow-band filters. Fields are C: COSMOS, U: UDS. Emitters are the narrow-band excess sources remaining after excluding artefacts and stars. Volumes are presented in units of 10$^4$\,Mpc$^3$ and correspond to the total volumes probed; deeper pointings cover relatively smaller volumes. SFRs (limiting SFRs) are presented uncorrected for dust extinction. The number of spectroscopically confirmed H$\alpha$ emitters are presented in the $z$ H$\alpha$ conf column, while the number of H$\alpha$ sources which are confirmed by an emission-line detection in another narrow-band filter are presented in Conf 2lines. Limiting SFRs are uncorrected for dust extinction and these represent the limit over the deepest field(s). Also, note that for all four redshifts the number of $z$-included sources missed by the selection is typically $<10$, while the $z$-rejected sources are typically $<10$ as well. Therefore, decrease of available spectroscopic redshifts with redshift does not introduce any bias. }
  \begin{tabular}{@{}ccccccccccccc@{}}
  \hline
  \bf Filter &   \bf Field &   \bf Detect &  \bf W/ Colours & \bf Emitters &  \bf Stars & \bf Artifacts  &  \bf H$\alpha$ & \bf $\bf z$ H$\alpha$ conf. & \bf Conf 2lines &  \bf Volume\,   &   \bf H$\alpha$ SFR \\
   NB & C/U & \bf 3\,$\bf \sigma$ & \# & \bf 3\,$\bf \Sigma$ & \# & \# &  \# & \# & \# & 10$^4$\,Mpc$^3$ & M$_\odot$\,yr$^{-1}$ \\
 \hline
   \noalign{\smallskip}
NB921 & C & 155542 & 148702 & 2819 & 247  & -- & 521 & 38 & -- & 5.1 & 0.03 \\
NB921 & U & 236718 & 198256 & 6957 & 775  & -- & 1221 & 8  & -- & 5.1 & 0.01 \\
 \hline
NB$_{\rm J}$ & C & 32345 & 31661 & 700 & 40 & 46 & 425 & 81 & 158  & 7.9 & 1.5  \\
NB$_{\rm J}$ & U & 21233 & 19916 & 551 & 49  & 30 & 212 & 14  & 79  & 11.1 & 1.5 \\
 \hline
NB$_{\rm H}$ & C & 65912 & 64453 & 723 & 60  & 63 & 327 & 28 & 158 & 49.1 & 3.0 \\
NB$_{\rm H}$ & U & 26084 & 23503 & 418 & 23  & 5 & 188 & 18 & 188 & 22.8 & 5.0  \\
 \hline
NB$_{\rm K}$ & C & 99395 & 98085 & 1359 & 78 & 56 & 588 & 4 & 125  & 54.8 & 5.0 \\
NB$_{\rm K}$ & U & 28276 & 26062 & 399 & 28 & 10 & 184 & 2 & 30  & 22.4 & 10.0  \\
H$_2$ & C & 1054 & 940 & 52 & 3 & 2 & 31 & 0 & 3 & 0.9 & 3.5 \\
H$_2$ & U & 1193 & 1059 &  33 & 7 & 1 & 14  & 0 & 0 & 0.9 & 3.5   \\
 \hline
\end{tabular}
\label{numbers}
\end{table*}

%
%
\begin{figure*}
\centering
\includegraphics[width=17.9cm]{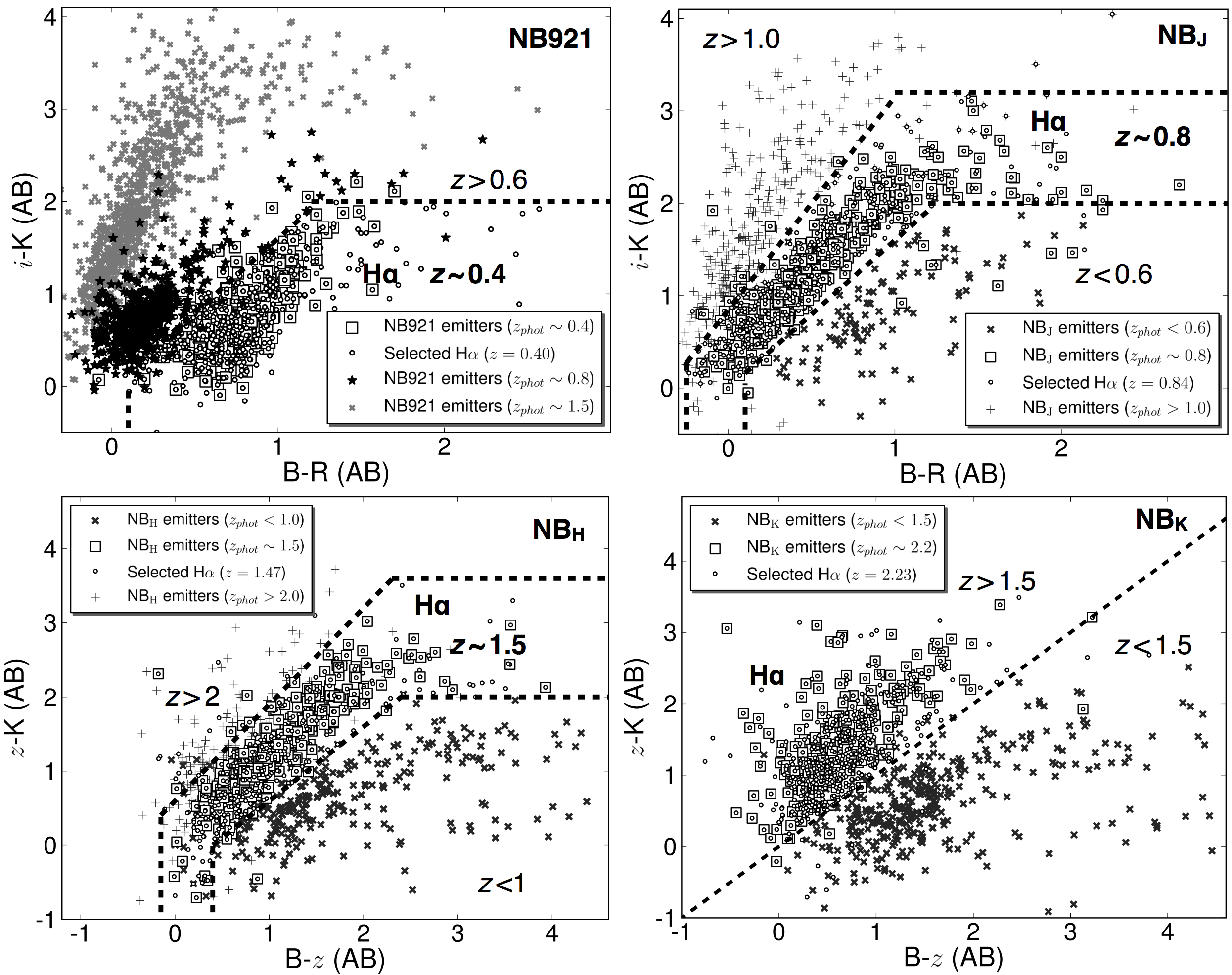}
\caption[COLOR-COLOR]{$Top$ $panels$: colour-colour separation of the different emitters in the various samples used to isolate H$\alpha$ emitters at the four different redshifts. Different symbols show narrow-band emitters with different photometric redshifts, and the final emitters selected to be H$\alpha$ (see Section 2.6). The $(B-R)$ vs. $(i-K)$ colour-colour separation is used to cleanly select $z=0.4$ H$\alpha$ emitters with the NB921 data, but also $z=0.84$ H$\alpha$ emitters. The $z=0.4$ selection points are [0.1,0.18] and [1.24,2.01], while the selection of $z=0.84$ H$\alpha$ emitters is obtained with the set of points: [$-$0.26,0.26], [0.1,0.18],[1.01,3.21] and [1.24,2.01] (see selection regions shown in the top panels of the Figure). $Bottom$ $panels$: the $(B-z)$ vs $(z-K)$ colours are used in different ways (to maximise completeness) to obtain first pass samples of potential $z=1.47$ H$\alpha$ emitters (defined by points: [$-$0.15,0.40], [0.41,0.0], [2.31,3.6] and [2.41,2.0]) and $z=2.23$ H$\alpha$ emitters \citep[][$z-$K$>$\,B$-z$]{Daddi04}. Note, however, that for the H$\alpha$ samples at $z=1.47$ and $z=2.23$ the B$z$K selection is not able to differentiate effectively between H$\alpha$ and higher redshift emitters which can significantly contaminate the samples (see photometric redshift distribution). The higher redshift emitters are filtered out of the samples using another set of colours -- see Figure \ref{BRU}.  \label{BRIKS}}
\end{figure*}

\subsection{Selecting H$\alpha$ emitters} \label{selecting_Ha}

Samples of H$\alpha$ emitters at the various redshifts are selected using a combination of broad-band colours (colour--colour selections) and photometric redshifts (when available). Colour--colour separation of emitters are different for each redshift, and for some redshifts two sets of colour--colour separations are used to reduce contamination to a minimum. Additionally, spectroscopically confirmed sources are included, and sources confirmed to be other emission lines removed from the samples -- but the reader should note that at all four redshifts the number of $z$-included sources missed by the selection is typically $<10$, while the $z$-rejected sources are typically $<10$ as well. Therefore, the decrease of available spectroscopic redshifts with redshift does not introduce any bias. 

Additionally, sources found to be line emitters in two (or three, for H$\alpha$ emitters at $z=2.23$) bands, making them robust H$\alpha$ candidates are also included in the samples, even if they have been missed by the colour--colour and photometric selection (although it is found that only very few real H$\alpha$ sources are missed by the selection criteria). Table \ref{numbers} provides the number of sources, including spectroscopically confirmed ones for each field at each redshift. Within the samples of narrow-band excess sources, 20 (NB921), 54 (NB$_{\rm J}$), 49 (NB$_{\rm H}$) and 47 (NB$_{\rm K}$) per cent are H$\alpha$ emitters at redshifts $z=0.4$, 0.84, 1.47 and 2.23, respectively.

\subsubsection{H$\alpha$ emitters at $z=0.4$} \label{selecting_Ha04}

The selection of H$\alpha$ emitters at $z=0.4$ is primarily done by selecting sources for which $0.35<z_{\rm phot}<0.45$. For further completeness, the $BRiK$ ($B-R$ vs $i-K$) colour-colour selection (see Figure \ref{BRIKS} and S12) is then applied to recover real H$\alpha$ sources without photometric redshifts. The selection method can then be accessed by using spectroscopic redshifts \citep[from $z$COSMOS; ][]{Lilly09}, which are available for 38 sources. Thirty six sources are confirmed to be at $z=0.391-0.412$, while 2 sources are [N{\sc ii}] emitters. This implies a very high completeness of the sample, and a contamination of $\sim5$ per cent over the entire sample. Contaminants have been removed, and spectroscopic sources added. A total of 1742 H$\alpha$ emitters at $z=0.4$ are selected.

\subsubsection{H$\alpha$ emitters at $z=0.84$} \label{selecting_Ha147}

Sources are selected to be H$\alpha$ emitters at $z=0.84$ if $0.75<z_{\rm phot}<0.95$ or if they satisfy the $BRiK$ (see Figure \ref{BRIKS}; S09) colour-colour selection for $z\sim0.8$ sources. Additionally, sources with $1.3<z_{\rm phot}<1.7$ (likely H$\beta$/[O{\sc iii}] $z\approx1.4$ emitters) and $2.0<z_{\rm phot}<2.5$ (likely $z=2.23$ [O{\sc ii}] emitters) are removed to further reduce contamination from higher redshift emitters. Sources with spectroscopically confirmed redshifts are included and sources with other spectroscopically confirmed lines are removed. In practice, 6 H$\alpha$ spectroscopic sources missed by the selection criteria are introduced in the sample; 7 sources found in the sample are not H$\alpha$ -- a mix of [S{\sc ii}], [N{\sc ii}] and [O{\sc iii}] emitters. A total of 95 sources are spectroscopically confirmed as H$\alpha$, while 237 sources are confirmed as dual H$\alpha$-[O{\sc iii}] emitters. A total of 637 H$\alpha$ emitters at $z=0.84$ are selected.

\subsubsection{H$\alpha$ emitters at $z=1.47$} \label{selecting_Ha147}

Note that, as described in S12, the NB$_{\rm H}$ filter can be combined with NB921 (probing the [O{\sc ii}] emission line), to provide very clean, complete surveys of $z=1.47$ line emitters, as the filter profiles are extremely well-matched for a dual H$\alpha$-[O{\sc ii}] survey. By applying the dual narrow-band selection, a total of 346 H$\alpha$-[O{\sc ii}] emitters are robustly identified in COSMOS and UDS. However, the dual narrow-band selection is only complete ($>98$\% complete) if the NB921 survey probes down to [O{\sc ii}]/H$\alpha$ $\sim0.1$ (c.f. S12), which is not the case for the deepest NB$_{\rm H}$ COSMOS coverage. Additionally, only the central 1.1\,deg$^2$ region of the COSMOS field has been targeted with the NB921 filter.

\begin{figure}
\centering
\includegraphics[width=7.8cm]{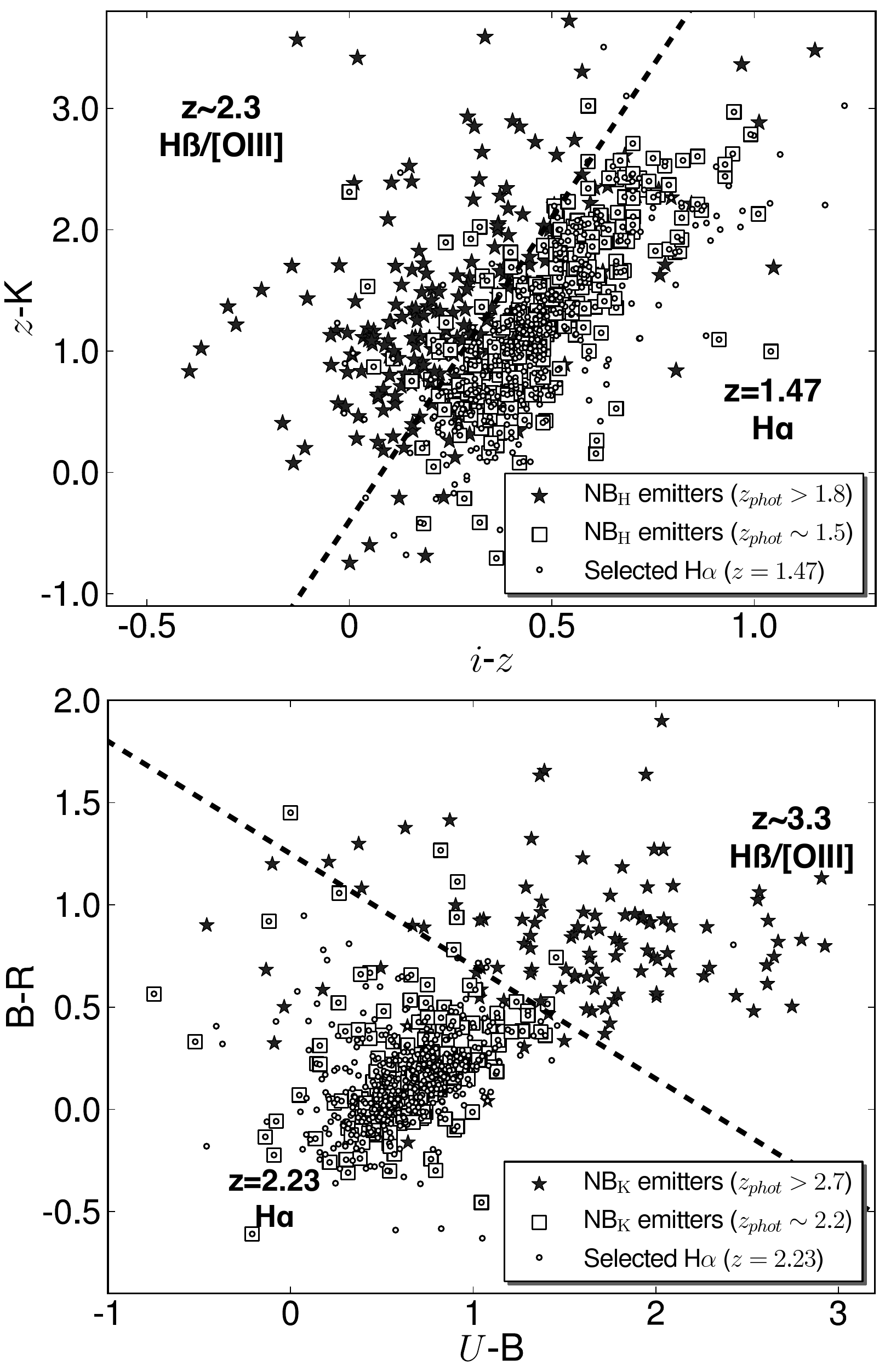}
\caption[COLOR-COLOR2]{$Top$: Colour-colour separation of $z=1.47$ H$\alpha$ emitters and those at higher redshift ($z\sim2.3$ H$\beta$/[O{\sc iii}], $z\sim3.3$ [O{\sc ii}]); separation is obtained by $(z-K)<5(i-z)-0.4$.  $Bottom$: colour-colour separation of $z\sim3$ H$\beta$/[O{\sc iii}] emitters from $z=2.23$ H$\alpha$ emitters, given by $(B-R)<-0.55(U-B)+1.25$. Using the B$z$K colour-colour separation only results in some contamination of the H$\alpha$ sample with higher redshift emitters -- the $B-R$ vs $U-B$ colour-colour separation allows to greatly reduce that. The Figure also indicates the location of each final H$\alpha$ selected source in the colour-colour plot. Note that sources shown as $z_{phot}\sim1.5$ and $z_{phot}\sim2.2$ are those with photometric redshifts which are within $\pm0.2$ of those values. \label{BRU}}
\end{figure}

In order to select H$\alpha$ emitters in areas where the NB921 is not deep enough to provide a complete selection, or where NB921 data are not available, the following steps are taken. Sources are selected if $1.35<z_{\rm phot}<1.55$, or if they satisfy the $z\sim1.5$ $BzK$ ($B-z$ vs. $z-K$) criteria defined in Figure \ref{BRIKS}, which is able to recover the bulk of the dual narrow-band emitters and sources with high quality photometric redshifts of $z\sim1.5$. However, the $z\sim1.5$ $BzK$ selection, although highly complete, is still contaminated by higher redshift emitters. In order to exclude likely higher redshift sources an additional $ziK$ ($i-z$ vs. $z-K$; see S12) colour-colour separation is used (see Figure \ref{BRU}), in combination with rejecting sources with $z_{\rm phot}>1.8$.

%
%

The selection leads to a total sample of 515 robust H$\alpha$ emitters at $z=1.47$, by far the largest sample of H$\alpha$ emitters at $z\sim1.5$. Comparing the double NB921 and NB$_{\rm H}$ analysis with the colour and photo-$z$ selection (for sources for which the NB921 data are deep enough to detect [O{\sc ii}]) shows that the colour and photo-$z$ selection by itself results in a contamination of $\approx15$ per cent, and a completeness of $\approx85$ per cent. However, as the double NB921 and NB$_{\rm H}$ analysis has been used wherever the data are available and sufficiently deep, the contamination of the entire sample is estimated to be lower ($\approx5$ per cent), and the completeness higher $(\approx95$ per cent).

%
%

\subsubsection{H$\alpha$ emitters at $z=2.23$} \label{selecting_Ha223}

As can be seen from the photometric redshift distribution in Figure \ref{photoz}, the high quality photo-$z$s in the COSMOS and UDS fields can provide a powerful tool to select $z=2.23$ sources. However, the sole use of the photometric redshifts can not result in clean, high completeness sample of $z=2.23$ H$\alpha$ emitters, not only because reliable photometric redshifts are not available for 35 per cent of the NB$_{\rm K}$ emitters, at the faint end, but also because the errors in the photometric redshifts will be much higher at $z\sim2.2$ than at lower redshift (particularly as one is selecting star-forming galaxies). Nevertheless, although spectroscopy only exists for a few H$\alpha$ $z=2.23$ sources ($z$COSMOS and UDS compilation), double line detections between NB$_{\rm K}$ and one of NB$_{\rm H}$ ([O{\sc iii}]) and/or NB$_{\rm J}$ ([O{\sc ii}]) allow the identification of 155 secure H$\alpha$ emitters. These can be used to optimise the selection criteria and estimate the completeness and contamination of the sample.

The selection of H$\alpha$ emitters is done in the same way for both COSMOS and UDS, and for both WFCAM and HAWK-I data. An initial sample of $z=2.23$ H$\alpha$ emitters is obtained by selecting sources for which $1.7<z_{\rm phot}<2.8$, where the limits were determined using the distribution of photometric redshifts found for confirmed H$\alpha$ emitters at $z=2.23$ (this selects 525 sources, of which 3 are spectroscopically confirmed to be contaminants and 87 are double/triple line emitters and thus robust $z=2.23$ H$\alpha$ emitters). Because some sources lack reliable photometric redshifts, the colour selection $(z-K)>(B-z)$ is used recover additional $z\sim2$ faint emitters. This colour-colour selection is a slightly modified version of the standard $BzK$ colour-colour separation \citep{Daddi04}\footnote{The selection was modified because the Daddi et al. cut was designed to select $z>1.4$ sources, while here $z=2.23$ emitters are targeted. The precise location of the new cut, which is 0.2 magnitudes higher/redder in $z-K$ than that of Daddi et al., is motivated by the confirmed H$\alpha$ emitters and by the need to minimise contamination from $z<2$ sources).}. It selects 274 additional H$\alpha$ candidates (and re-selects 90\% of those selected through photometric redshifts), and guarantees a high completeness of the H$\alpha$ sample (see Figure \ref{BRIKS}). However, the $BzK$ selection also selects $z\sim3.3$ H$\beta$/[O{\sc iii}] emitters very effectively, and the contamination by such emitters needs to be minimised. In order to do this, sources with $z_{\rm phot}>3.0$ are excluded (121 sources). For sources for which a photometric redshift does not exist, a rest-frame UV colour-colour separation is used ($B-R$ vs. $U-B$; see Figure \ref{BRU}, probing the rest-frame UV), capable of broadly separating $z=2.23$ and $z\sim3.3$ emitters due to their different UV colours (see Figure \ref{BRU}; this removes a further 27 sources). Three further sources are removed as they are confirmed contaminants (Pa$\beta$, [S{\sc iii}] and [O{\sc iii}] at $z=0.65$, $z=1.23$ and $z=3.23$, respectively).

Overall, the selection leads to a total sample of 807 H$\alpha$ emitters, by far the largest sample of $z=2.23$ H$\alpha$ emitters ever obtained, and an order of magnitude larger than the previous largest samples presented by \cite{G08} and \cite{Hayes}. With the limited spectroscopy available, it is difficult to accurately determine the completeness and contamination of the sample, but based on the double/triple-line detections (155) and the confirmed contaminants which have been removed (6), the completeness is estimated to be $>90$ per cent, and contamination is likely to be $<10$ per cent.

\section{Analysis and Results: H$\alpha$ LF over 11 Gyrs } \label{LFs}

\subsection{Removing the contamination by the [N{\sc ii}] line}  \label{cont_adj_lines}

Due to the width of all filters in detecting the H$\alpha$ line, the adjacent [N{\sc ii}] lines can also be detected when the H$\alpha$ line is detected at the peak transmission of the filter. A correction for the [N{\sc ii}] line contamination is therefore done, following the relation given in S12. The relation has been derived to reproduce the full SDSS relation between the average $\log$([N{\sc ii}]/H$\alpha)$, $f$, and $\log$[EW$_0$([N{\sc ii}]+H$\alpha$)], $E$: $f=-0.924+4.802E-8.892E^2+6.701E^3-2.27E^4+0.279E^5$. This relation is used to correct all H$\alpha$ fluxes at $z=0.4$, $0.84$, $1.47$ and $2.23$. The median correction (the median [N{\sc ii}]/([N{\sc ii}]+H$\alpha$)) is $\approx0.25$.

%
%
\begin{figure}
\centering
\includegraphics[width=8.2cm]{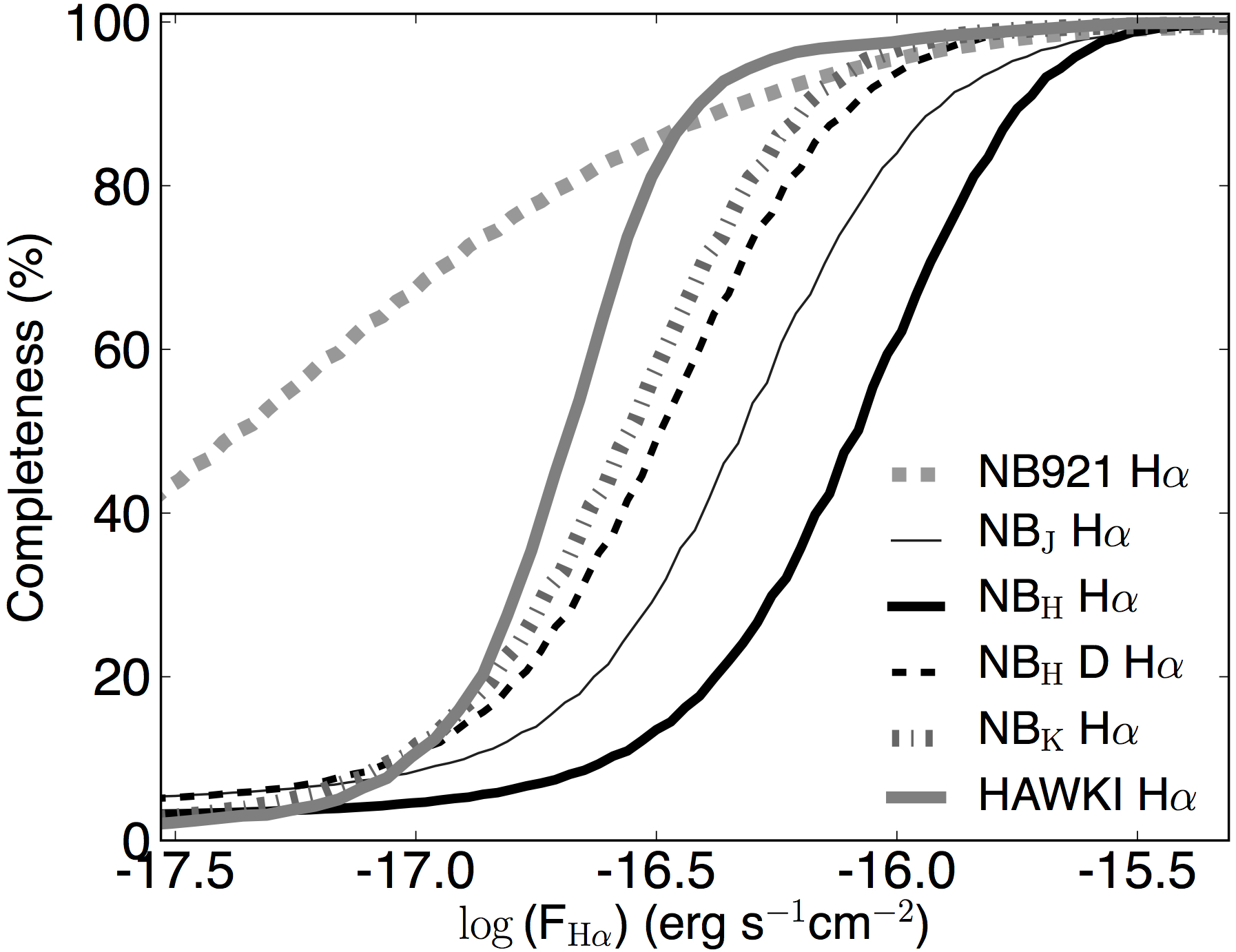}
\caption[Incompleteness]{Average completeness of the various narrow-band surveys as a function of H$\alpha$ flux. Note that the completeness of individual fields/frames for each band can vary significantly due to the survey strategy (e.g. see difference between one of the deep pointings in COSMOS, NB$_{\rm H}$ D and the average over all NB$_{\rm H}$ fields), and thus the completeness corrections are computed for each individual fields. \label{line_frac_fluxes}}
\end{figure}

\subsection{Completeness corrections: detection and selection}  \label{complet}

It is fundamental to understand how complete the samples are as a function of line flux. This is done using simulations, as described in S09 and further detailed in S12. The simulations consider two major components driving the incompleteness: i) the detection completeness (which depends on the actual imaging depth and the apertures used) and ii) the incompleteness resulting from the selection (both EW and colour significance). 

The detection completeness is estimated by placing sources with a given magnitude at random positions on each individual narrow-band image, and studying the recovery rate as a function of the magnitude of the source. For the large Subaru frames, 2500 sources are added for each magnitude, for WFCAM images 500, and for HAWK-I frames 100 sources are added for each realisation.

The individual line completeness estimates are performed in the same way for the data at the four different redshifts. A set of galaxies is defined, which is consistent with being at the approximate redshift (applying the same photometric redshift + colour-colour selections to all NB detected sources with no significant excess) but not having emission lines above the detection limit. Emission lines are then added to the sources, and the study of the recovery fraction is undertaken. The average completeness corrections as a function of H$\alpha$ flux are presented in Figure \ref{line_frac_fluxes}. Note that the simulations include the different EW/colour cuts used in selecting line emitters in all bands, and therefore take the EW limits and colour selection into account. Also note that because of the very different distributions of magnitudes of H$\alpha$ emitters from low to higher redshift, the EW/colour cut is a much more important source of incompleteness for low redshift H$\alpha$ emitters than for the highest redshift, $z=2.23$.

It should be noted that because of the differences in depth, simulations are conducted for each individual frame, and the appropriate completeness corrections applied accordingly when computing the luminosity function. For any given completeness correction applied, an uncertainty of 20\% of the size of the applied correction is added in quadrature to the other uncertainties to account for the uncertainties in deriving such corrections.

\subsection{Volume}  \label{volume}

At $z=0.4$, the total area surveyed is 1.68 deg\,$^2$. The NB921 filter, centred at 9196\,\AA \ and with a FWHM of 132\,\AA \ can probe the H$\alpha$ line (using the top hat approximation) from $z_{\rm min}=0.3907$ to $z_{\rm max}=0.4108$. This means that the narrow-band filter surveys an H$\alpha$ volume of $5.1\times10^4$\,Mpc$^3$\,deg$^{-2}$. The H$\alpha$ survey therefore probes a total volume of $8.8\times10^4$\,Mpc$^3$.

The NB$_{\rm J}$ filter (FWHM of 140\,\AA) can be approximated by a top hat, probing $z_{\rm min}=0.8346$ to $z_{\rm max}=0.8559$ for H$\alpha$ line detections, resulting in surveying $1.5\times10^5$\,Mpc$^3$\,deg$^{-1}$. As the total survey has covered 1.3\,deg$^2$, it results in a total volume of $1.9\times10^5$\,Mpc$^3$. Assuming the top hat (TH) model for the NB$_{\rm H}$ filter (FWHM of 211.1\,\AA, with $\lambda^{TH}_{min}=1.606\,\umu$m and $\lambda^{TH}_{max}=1.627\,\umu$m), the H$\alpha$ survey probes a (co-moving) volume of $3.3\times10^5$\,Mpc$^3$\,deg$^{-2}$. Volumes are computed on a field by field basis as each field reaches a different depth (although the difference in volume is only important at the faintest fluxes). The total volume of the survey is $7.4\times10^5$\,Mpc$^3$. The volume down to the deepest depth is $3.9\times10^4$\,Mpc$^3$ (see Table 4 for details). The NB$_{\rm K}$ filter is centred on $\lambda=2.121$\,$\umu$m, with a FWHM of 210\,\AA. Using the top hat approximation for the filter, it can probe the H$\alpha$ emission line from $z_{\rm min}=2.2147$ to $z_{\rm max}=2.2467$, so with a $\Delta z=0.016$. The H$_2$ filter therefore probes a volume of $3.8\times10^5$\,Mpc$^3$\,deg$^{-2}$.

The HAWK-I survey uses a slightly different H$_2$ filter, centred on $\lambda=2.125$\,$\umu$m, with FWHM\,$=300$\,\AA. A top hat is an even better approximation of the filter profile, with $z_{\rm min}=2.2139$ to $z_{\rm max}=2.2596$ for H$\alpha$ line detections. The filter effectively probes $5.5\times10^5$\,Mpc$^3$\,deg$^{-2}$. Each HAWK-I pointing covers only about 13.08\,arcmin$^2$, and so the complete HAWK-I survey (COSMOS and UDS, 0.0156\,deg$^2$) probes a total volume of $1.7\times10^4$\,Mpc$^3$. Note that the survey conducted by \cite{Hayes} (using a narrower NB filter), although deeper, only probed $5.0\times10^3$\,Mpc$^3$, so a factor of 3 smaller in volume and over a single field. Table 4 presents a summary of the volumes probed as a function of H$\alpha$ luminosity and the number of sources detected at each redshift.

\subsection{Filter Profiles: volume corrections}  \label{filter_profiles}

None of the narrow band filters  are perfect top hats (see Figure 1). In order to model the effect of this bias on estimating the volume (luminous emitters will be detectable over larger volumes -- although, if seen in the filter wings, they will be detected as fainter emitters), a series of simulations is done, following S09 and S12. Briefly, a top hat volume selection is used to compute a first-pass (input) luminosity function and derive the best fit. The fit is used to generate a population of simulated H$\alpha$ emitters (assuming they are distributed uniformly across redshift); these are then folded through the true filter profile, from which a recovered luminosity function is determined. Studying the difference between the input and recovered luminosity functions shows that the number of bright emitters is underestimated, while faint emitters can be slightly overestimated (c.f. S09 for details), but the actual corrections are different for each filter and each input luminosity function. This allows correction factors to be estimated -- these are then used to obtain the corrected luminosity function. Corrections are computed for each individual narrow-band filter.

\subsection{Extinction Correction}  \label{ext_corr}

The H$\alpha$ emission line is not immune to dust extinction. Measuring the extinction for each source can in principle be done by several methods, one of which is the comparison between H$\alpha$ and far-infrared determined SFRs (see Ibar et al. in prep.), while the spectroscopic analysis of Balmer decrements also provides a very good estimate of the extinction. As shown in S12, the median [O{\sc ii}]/H$\alpha$ line ratio of a large sample of galaxies can also be reasonably well calibrated (using Balmer decrement) as a dust extinction indicator (see Sobral et al. 2012 for more details). For the COSMOS $z=1.47$ sample, this results in $A_{\rm H\alpha}=0.8$\,mag (although there is a bias towards lower extinction due to the fact that the NB921 survey is not deep enough to recover sources with much higher extinctions). However, for UDS (where a sufficiently deep NB921 coverage is available) a A$_{\rm H\alpha}\approx1$\,mag of extinction at H$\alpha$ is shown to be an appropriate median correction at $z=1.47$ (see S12). That is also similar has been found at $z=0.84$  \citep[][$A_{\rm H\alpha}\approx1.2$]{Garn2010a}. The dependence of extinction on observed luminosity is also relatively small (S12) at $z\sim1.5$ -- therefore, for simplicity and for an easier comparison, a simple 1 mag of extinction is applied for the four redshifts and for all observed luminosities.

Note that S12 still find a relatively mild luminosity dependence, but one which is offset to the local Universe relation \citep[e.g.][]{Hopkins} by 0.5 mag in $A_{\rm H\alpha}$. Nevertheless, one could interpret this differently, as a single relation that holds at both $z\sim1.5$ and $z\sim0$, provided that luminosities at both $z\sim0$ and $z\sim1.5$ are divided by $L^*_{\rm H\alpha}$ at the corresponding epochs; this would imply that the typical extinction does not depend on SFR or H$\alpha$ luminosity in an absolute manner, but rather that it depends on how star-forming or luminous a source is relative the normal star-forming galaxy at that epoch.

\subsection{H$\alpha$ Luminosity Functions at $\bf z=0.40,0.84,1.47,2.23$}  \label{LF_Ha}

%
%
\begin{figure*}
\centering
\includegraphics[width=14.8cm]{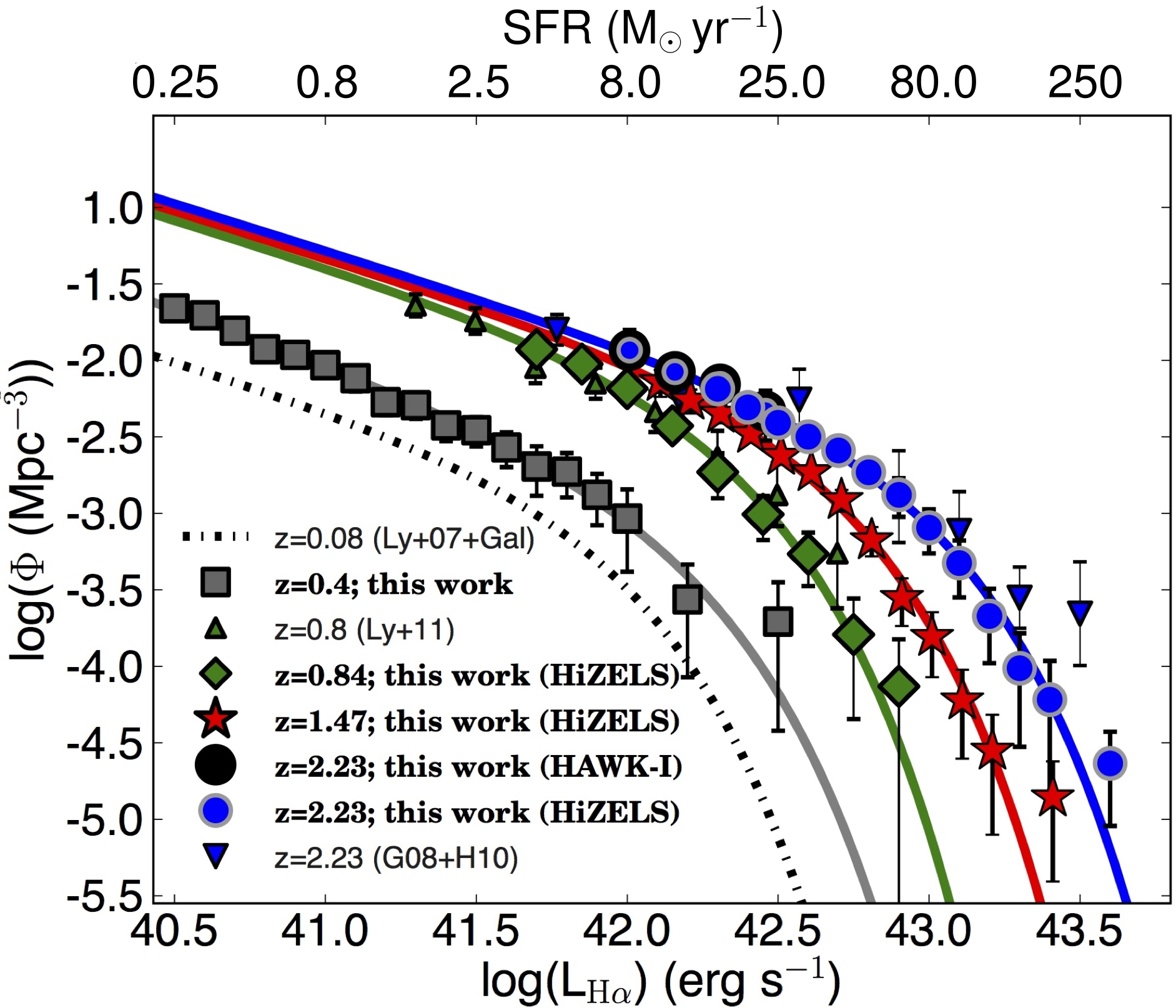}
\caption[The H$\alpha$ luminosity function evolution]{The H$\alpha$ luminosity function evolution revealed by deep and wide narrow-band surveys at $z=0.4$, $z=0.84$, $z=1.47$ and $z=2.23$ presented in this work. The data are combined and compared to other studies \citep{Gallego95,Ly2007,G08,Hayes}, and clearly confirm the strong evolution of the luminosity function in the last $\sim11$\,Gyrs, mostly through a continuous increase in L$_{\rm H\alpha}^*$ from $z=0$ to $z=2.23$. Note that the $z=0.4$ H$\alpha$ luminosity function is constrained down to even lower luminosities which are not shown in the figure. Also, note that all H$\alpha$ luminosities have been corrected by extinction with A$_{\rm H\alpha}=1$\,mag, and that SFRs shown are based on that extinction correction. SFRs derived directly from observed H$\alpha$ luminosities are a factor of 2.5 lower and H$\alpha$ luminosities uncorrected for extinction are 0.4 dex lower.\label{LF_Halpha_HALPHA}}
\end{figure*}

By taking all H$\alpha$ selected emitters at the four different redshifts, the H$\alpha$ luminosity function is computed at 4 very different cosmic times, reaching a common observed luminosity limit of $\approx10^{41.6}$\,erg\,s$^{-1}$ for the first time in a consistent way over $\sim11$\,Gyrs. As previously described, the method of S09 and S12 is applied to correct for the real profile (see Section \ref{filter_profiles}). Candidate H$\alpha$ emitters are assumed to be at $z=0.4$, $0.84$, $1.47$ and $2.23$ for luminosity distance calculations. Results can be found in Figure \ref{LF_Halpha_HALPHA} and Table \ref{LF_NUMBERS}. Errors are Poissonian, but they include a further 20\% of the total completeness corrections added in quadrature.

All derived luminosity functions are fitted with Schechter functions defined by three parameters $\alpha$, $\phi ^*$ and $L^*$:

\begin{equation}
  \phi(L) \rm dL = \it \phi^* \left(\frac{L}{L^*}\right)^{\alpha} e^{-(L/L^*)} \rm d\it\left(\frac{L}{L^*}\right),
\end{equation}
which are found to provide good fits to the data at all redshifts. In the $\log$ form, the Schechter function is given by:

\begin{equation}
  \phi(L) \rm dL = \ln10 \, \it \phi^* \left(\frac{L}{L^*}\right)^{\alpha} e^{-(L/L^*)} \left(\frac{L}{L^*}\right)\rm d\log L.
\end{equation}
Schechter functions are fitted to each luminosity function. The best fits for the H$\alpha$ luminosity functions at $z=0.4-2.23$ are presented in Table \ref{lfs__}, together with the uncertainties on the parameters (1\,$\sigma$). Uncertainties are obtained from either the 1$\sigma$ deviation from the best-fit, or the 1$\sigma$ variance of fits, obtained with a suite of multiple luminosity functions with different binning -- whichever is higher (although they are typically comparable). The best-fit functions and their errors are also shown in Figure \ref{LF_Halpha_HALPHA}, together with the $z\approx0$ luminosity function determined by \cite{Ly2007} -- which has extended the work by \cite{Gallego95} at $z\approx0$, for a local-Universe comparison. Deeper data from the literature are also presented for comparison; \cite{CHU11} for $z=0.8$, and \cite{Hayes} for $z=2.23$, after applying the small corrections to ensure the extinction corrections are consistent\footnote{The correction is applied to obtain data points corrected for extinction by 1 mag at H$\alpha$.}.

%
%
\begin{figure*}
\centering
\includegraphics[width=17.cm]{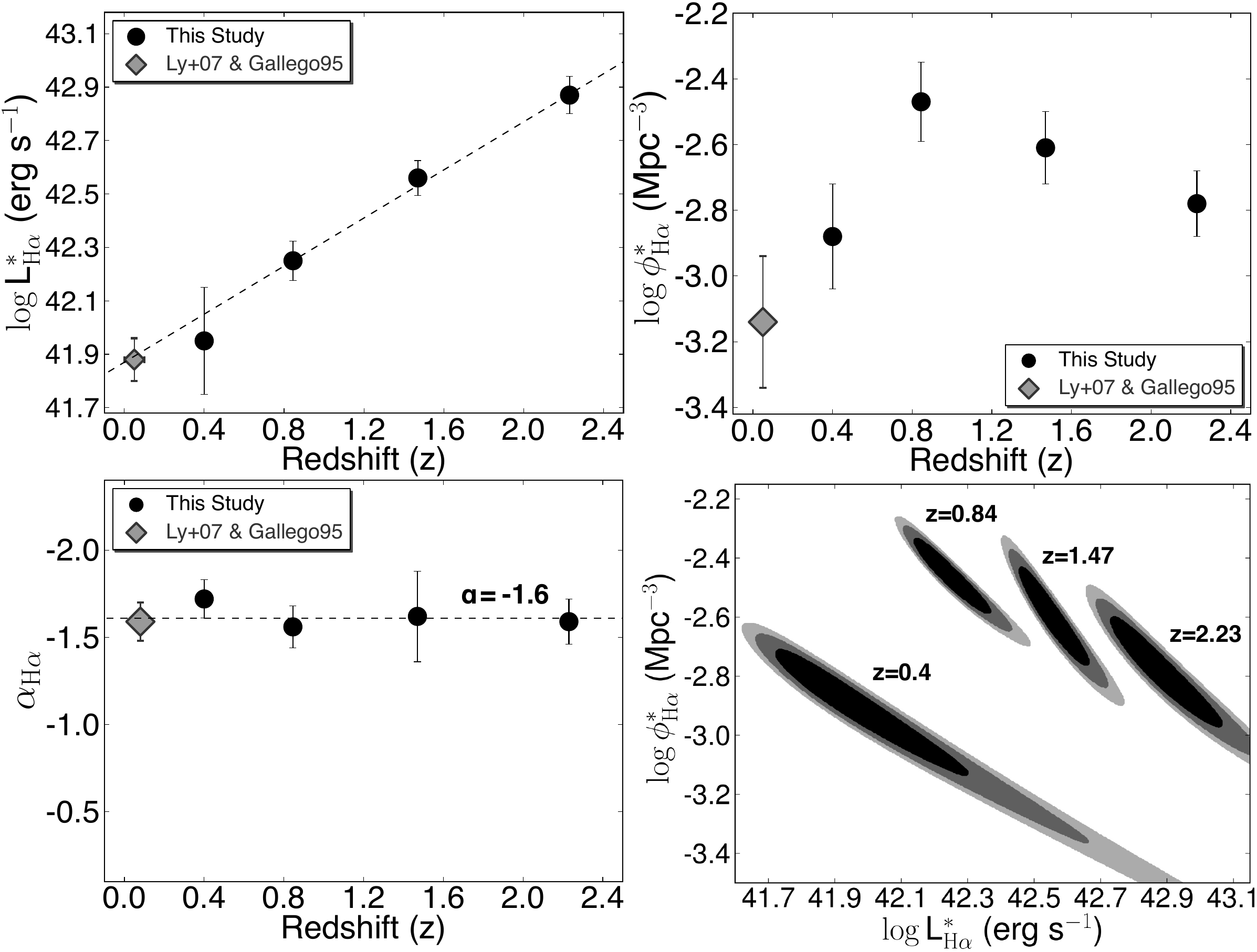}
\caption[LF EVO]{The evolution of the Schechter function parameters which best fit the H$\alpha$ luminosity function since $z=2.23$. $Top$ $left$: the evolution of L$^*_{\rm H\alpha}$ as a function of redshift, revealing that the break of the luminosity function evolves significantly from $z=0$ to $z=2.23$ by a factor of 10, which can be simply parameterised by $\log\,L^*_{\rm H\alpha}=0.45z+41.87$. $Top$ $right$: the evolution of $\phi^*$, which seems to rise mildly up to $z\sim1$ and decrease again up to $z\sim2.2$. $Bottom$ $left$: the faint end slope, however, is not found to evolve at all from $z=0.0$ to $z=2.23$ within the scatter and the errors, pointing towards $\alpha=-1.60\pm0.08$ for the faint-end slope of the H$\alpha$ luminosity function across the last 11 Gyrs. $Bottom$ $right$: The 1, 2 and 3\,$\sigma$ contours of the best fits to the combination of L$^*_{\rm H\alpha}$ and $\phi^*$ fixing $\alpha=-1.6$. \label{evo_PARAMS}}
\end{figure*}

The results not only reveal a very clear evolution of the H$\alpha$ luminosity function from $z=0$ to $z=2.23$, but they also allow for a detailed investigation of exactly how the evolution occurs, in steps of $\sim2-3$\,Gyrs. The strongest evolutionary feature is the increase in $L_{\rm H\alpha}^*$ as a function of redshift from $z=0$ to $z=2.23$ (see Figure \ref{evo_PARAMS}), with the typical H$\alpha$ luminosity at $z\sim2$ ($L_{\rm H\alpha}^*$) being 10 times higher than locally. This is clearly demonstrated in Figure \ref{evo_PARAMS}, which shows the evolution of the Schechter function parameters describing the H$\alpha$ luminosity function. The L$_{\rm H\alpha}^*$ evolution from $z\sim0$ to $z\sim2.2$, can be simply approximated as $\log\,L^*=0.45z+\log\,L^*_{z=0}$, with $\log\,L^*_{z=0}=41.87$ (see Figure \ref{evo_PARAMS}). At the very bright end ($L>4L^*$), and particularly at $z>1$, there seems to be a deviation from a Schechter function. Follow-up spectroscopy of such luminous H$\alpha$ sources has recently been obtained for a subset of the $z=1.47$ sample, and unveil a significant fraction of narrow and broad-line AGN (with strong [N{\sc ii}] lines as well) which become dominant at the highest luminosities (Sobral et al. in prep). It is therefore likely that the deviation from a Schechter function is being mostly driven by the increase in the AGN activity fraction at such luminosities, particularly due to the detection of rare broad-line AGN and from very strong [N{\sc ii}] emission.

The normalisation of the H$\alpha$ luminosity function, $\phi^*$, is also found to evolve, but much more mildly. There is an increase of $\phi^*$ up to $z\sim1$ (by a factor $\sim4$)\footnote{Note that the difference in $\phi^*$ to the Sobral et al. (2009) H$\alpha$ luminosity function is mostly driven by $\phi^*_{S09}$ reported there being $\phi^*_{S09}=\phi^*\times \ln10$ (due to the fitting to dLogL without taking the ln10 factor into account -- see Sobral et al. 2012), which accounts for a factor $\approx2.23$. The remaining difference (a factor $\sim1.5$) is an actual difference driven by the improved data reduction, selection of emitters (3$\Sigma$ instead of 2.5$\Sigma$), completeness and cleanness of the catalogues of H$\alpha$ emitters (particularly due to the significantly improved photometric redshifts and a larger number of spectroscopic redshifts). }, and then this decreases again for higher redshifts by a factor of $\sim2$ from $z\sim1$ to $z=2.23$ -- see Figure \ref{evo_PARAMS}. By fitting a simple quadratic model to describe the data, one finds that the parametrisation: $\log\phi^*=-0.38z+z-3.18$ provides a good fit for $z=0-2.23$, but the current data can only exclude a model with a constant $\phi^*$ at a $<2$\,$\sigma$ level. The statistical significance for evolution in $\phi^*$ becomes even lower ($<1$\,$\sigma$) if one restricts the analysis to $z=0.4-2.23$.

The faint-end slope, $\alpha$, is found to be relatively steep from $z\sim0$ up to $z=2.23$ (when compared to a canonical $\alpha=-1.35$), and it is not found to evolve. The median $\alpha$ over $0<z<2.23$ is $-1.60\pm0.08$. Very deep data from \cite{Hayes} and \cite{CHU11} not only agree well with such faint-end slope, but even more importantly, their data at the faintest luminosities are also very well fitted by the best-fit $z=2.23$ and $z=0.84$ luminosity functions. If those data points are included and used to re-fit the luminosity functions at those 2 redshifts, the resulting best-fit faint-end slopes remain the same, but the error in $\alpha$ is reduced by $\sim10-15$ per cent.

It is therefore shown that by measuring the H$\alpha$ luminosity function in a consistent way, and using multiple fields, the faint-end slope can be very well approximated by a constant $\alpha=-1.6$ at least up to $z=2.23$. This shows that while the faint-end slope truly is steep at $z\sim2$, it does not become significantly steeper from $z\sim0$ to $z\sim2$, and rather has remained relatively constant for the last 11\,Gyrs (our data can not rule out weak evolution). The potential strong steepening of the faint-end slope, which has been previously reported \citep[e.g.][]{Hayes} may in part be a result of comparing different data-sets which probe different ranges in luminosity, use different completeness corrections, different selection of emitters and probe a different parameter space. Furthermore, the results from \cite{SOBRAL10B} show that the faint-end slope depends relatively strongly on environment ($\alpha\sim-1.1$ for the densest clusters to $\alpha\sim-1.9$ for the poorest regions), which indicates that the changes in the faint-end slope measured before may also have resulted by the relatively small areas which can (by chance) probe different environments. Note that this is not the case for this paper because the multi-epoch H$\alpha$ surveys cover $\sim2$\,deg$^2$ areas over two independent fields and are able to cover a wide range of environments. Indeed, apart from the rich, dense structures presented in \cite{SOBRAL10B} at $z=0.84$, our the H$\alpha$ survey is also able to probe significantly overdense regions even at $z=2.23$ \citep[see][for details on a significant H$\alpha$-detected overdensity in the COSMOS field]{GEACH12}. By splitting the sample in a similar way to \cite{SOBRAL10B} (isolating the over density in COSMOS and nearby regions), a variation of $\alpha$ with local density is clearly recovered, consistent with the results at $z=0.84$, i.e., overdense regions present a much shallower $\alpha$ ($\sim-1.3$), while the general field regions have a steeper ($\alpha\sim-1.7$) faint-end slope. The dependence of $\alpha$ on environment since $z=2.23$ will be carefully quantified in a forthcoming paper.

%
%
\begin{table}
 \centering
  \caption{Luminosity Functions from HiZELS. L$_{\rm \bf H\alpha}$ has been corrected for both [N{\sc ii}] contamination and for dust extinction (using A$_{\rm H\alpha}=1$\,mag).  Volumes assuming top hat filters. $\phi$ corr has been corrected for both incompleteness and the fact that the filter profile is not a perfect top hat.}
  \begin{tabular}{@{}ccccc@{}}
  \hline
  \bf $\bf \log$\bf L$_{\rm \bf H\alpha}$ & \bf Sources &$\bf \phi$ \bf obs & $\bf \phi$ \bf corr & \bf Volume  \\
 \hline
 \bf  $\bf z=0.40$ & \# & Mpc$^{-3}$ & Mpc$^{-3}$ & 10$^4$\,Mpc$^3$ \\
  \hline
   \noalign{\smallskip}
$40.50\pm0.05$  & $128$ & $-1.84\pm0.04$ & $-1.66\pm0.04$ & 8.8  \\
$40.60\pm0.05$  & $147$ & $-1.78\pm0.04$ & $-1.70\pm0.04$ & 8.8  \\
$40.70\pm0.05$  & $118$ & $-1.87\pm0.04$ & $-1.81\pm0.04$ & 8.8  \\
$40.80\pm0.05$  & $86$ & $-2.01\pm0.05$ & $-1.93\pm0.05$ & 8.8  \\
$40.90\pm0.05$  & $56$ & $-2.20\pm0.06$ & $-1.96\pm0.07$ & 8.8 \\
$41.00\pm0.05$  & $54$ & $-2.21\pm0.06$ & $-2.03\pm0.07$ & 8.8  \\
$41.10\pm0.05$  & $34$ & $-2.41\pm0.08$ & $-2.12\pm0.09$ & 8.8  \\
$41.20\pm0.05$  & $36$ & $-2.39\pm0.08$ & $-2.27\pm0.08$ & 8.8  \\
$41.30\pm0.05$  & $33$ & $-2.43\pm0.08$ & $-2.29\pm0.09$ & 8.8  \\
$41.40\pm0.05$  & $25$ & $-2.55\pm0.10$ & $-2.42\pm0.10$ & 8.8  \\
$41.50\pm0.05$  & $25$ & $-2.55\pm0.10$ & $-2.46\pm0.11$ & 8.8  \\
$41.60\pm0.05$  & $17$ & $-2.71\pm0.12$ & $-2.57\pm0.13$ & 8.8  \\
$41.70\pm0.05$  & $10$ & $-2.94\pm0.17$ & $-2.69\pm0.19$ & 8.8  \\
$41.80\pm0.05$  & $11$ & $-2.90\pm0.16$ & $-2.73\pm0.17$ & 8.8  \\
$41.90\pm0.05$  & $8$ & $-3.04\pm0.19$ & $-2.88\pm0.20$ & 8.8  \\
$42.00\pm0.05$  & $4$ & $-3.34\pm0.30$ & $-3.03\pm0.35$ & 8.8  \\
$42.20\pm0.10$  & $3$ & $-3.45\pm0.36$ & $-3.56\pm0.51$ & 8.8  \\
$42.50\pm0.15$  & $2$ & $-3.64\pm0.53$ & $-3.71\pm0.71$ & 8.8  \\
 \hline
 \bf $\bf z=0.84$& \# & Mpc$^{-3}$ & Mpc$^{-3}$ & 10$^4$\,Mpc$^3$ \\
  \hline
$41.70\pm0.075$  & $218$ & $-2.12\pm0.03$ & $-1.93\pm0.03$ & 19.1  \\
$41.85\pm0.075$  & $222$ & $-2.11\pm0.03$ & $-2.02\pm0.03$ & 19.1  \\
$42.00\pm0.075$  & $107$ & $-2.43\pm0.04$ & $-2.18\pm0.04$ & 19.1  \\
$42.15\pm0.075$  & $54$ & $-2.72\pm0.06$ & $-2.43\pm0.06$ & 19.1  \\
$42.30\pm0.075$  & $12$ & $-3.38\pm0.15$ & $-2.73\pm0.17$ & 19.1  \\
$42.45\pm0.075$  & $10$ & $-3.46\pm0.17$ & $-3.01\pm0.17$ & 19.1  \\
$42.60\pm0.075$  & $7$ & $-3.61\pm0.21$ & $-3.27\pm0.21$ & 19.1  \\
$42.75\pm0.075$  & $2$ & $-4.16\pm0.53$ & $-3.79\pm0.55$ & 19.1  \\
$42.90\pm0.075$  & $1$ & $-4.46\pm0.90$ & $-4.13\pm1.51$ & 19.1  \\
   \hline
 \bf $\bf z=1.47$ & \# & Mpc$^{-3}$ & Mpc$^{-3}$ &10$^4$\,Mpc$^3$ \\
  \hline
$42.10\pm0.05$  & $25$ & $-2.20\pm0.10$ & $-2.13\pm0.10$ & 4.0  \\
$42.20\pm0.05$  & $32$ & $-2.37\pm0.08$ & $-2.25\pm0.09$ & 7.5  \\
$42.30\pm0.05$  & $62$ & $-2.55\pm0.06$ & $-2.34\pm0.06$ & 22.1  \\
$42.40\pm0.05$  & $86$ & $-2.67\pm0.05$ & $-2.47\pm0.05$ & 40.2  \\
$42.50\pm0.05$  & $101$ & $-2.78\pm0.05$ & $-2.62\pm0.05$ & 60.4  \\
$42.60\pm0.05$  & $106$ & $-2.83\pm0.04$ & $-2.73\pm0.04$ & 71.4  \\
$42.70\pm0.05$  & $43$ & $-3.23\pm0.07$ & $-2.91\pm0.08$ & 73.6  \\
$42.80\pm0.05$  & $23$ & $-3.50\pm0.10$ & $-3.18\pm0.11$ & 73.6  \\
$42.90\pm0.05$  & $9$ & $-3.91\pm0.18$ & $-3.55\pm0.18$ & 73.6  \\
$43.00\pm0.05$  & $5$ & $-4.17\pm0.26$ & $-3.81\pm0.26$ & 73.6  \\
$43.10\pm0.05$  & $3$ & $-4.39\pm0.37$ & $-4.22\pm0.38$ & 73.6  \\
$43.20\pm0.05$  & $2$ & $-4.57\pm0.53$ & $-4.55\pm0.55$ & 73.6  \\
$43.40\pm0.15$  & $2$ & $-4.57\pm0.53$ & $-4.86\pm0.55$ & 73.6  \\
   \hline
 \bf $\bf z=2.23$ & \# & Mpc$^{-3}$ & Mpc$^{-3}$ & 10$^4$\,Mpc$^3$ \\
  \hline
$42.00\pm0.075$  & $8$ & $-2.18\pm0.19$ & $-1.93\pm0.19$ & 0.8  \\ 
$42.15\pm0.075$  & $11$ & $-2.34\pm0.16$ & $-2.07\pm0.16$ & 1.6  \\ 
$42.30\pm0.05$  & $47$ & $-2.24\pm0.07$ & $-2.19\pm0.07$ & 6.7  \\ 
$42.40\pm0.05$  & $91$ & $-2.36\pm0.05$ & $-2.31\pm0.05$ & 20.9  \\ 
$42.50\pm0.05$  & $107$ & $-2.48\pm0.04$ & $-2.41\pm0.05$ & 32.7  \\
$42.60\pm0.05$  & $158$ & $-2.60\pm0.04$ & $-2.50\pm0.04$ & 63.3  \\
$42.70\pm0.05$  & $163$ & $-2.68\pm0.04$ & $-2.59\pm0.05$ & 77.2  \\
$42.80\pm0.05$  & $100$ & $-2.89\pm0.05$ & $-2.73\pm0.06$ & 77.2  \\
$42.90\pm0.05$  & $51$ & $-3.18\pm0.07$ & $-2.88\pm0.14$ & 77.2  \\
$43.00\pm0.05$  & $30$ & $-3.41\pm0.09$ & $-3.09\pm0.17$ & 77.2  \\
$43.10\pm0.05$  & $16$ & $-3.68\pm0.12$ & $-3.33\pm0.22$ & 77.2  \\
$43.20\pm0.05$  & $7$ & $-4.04\pm0.21$ & $-3.67\pm0.31$ & 77.2  \\
$43.30\pm0.05$  & $3$ & $-4.41\pm0.37$ & $-4.01\pm0.51$ & 77.2  \\
$43.40\pm0.05$  & $2$ & $-4.59\pm0.53$ & $-4.22\pm0.68$ & 77.2  \\
$43.60\pm0.15$  & $3$ & $-4.41\pm0.37$ & $-4.63\pm0.41$ & 77.2  \\
 \hline
\end{tabular}
\label{LF_NUMBERS}
\end{table}

The steep faint-end slope of the H$\alpha$ luminosity function is in very good agreement with the UV luminosity function at $z\sim2$ and above, and particularly consistent with a relatively non-evolving $\alpha\approx-1.6$. This can be seen by comparing the results in this paper with those presented by \cite{TREYER98}, \cite{Arnouts05}, and more recently, \cite{Oesch10}. It is also likely that (similarly to the H$\alpha$ luminosity function) the large scatter, and the different selections/corrections applied, have driven studies to assume/argue for a steepening of the UV luminosity faint-end slope, just like for the H$\alpha$ luminosity function -- see \cite{Oesch10}.

Overall, the results imply that the bulk of the evolution of the star-forming population from $z=0$ to $z\sim2.2$ is occurring as a strong boost in luminosity of all galaxies. The UV luminosity results also show very similar trends to the H$\alpha$ LF, by revealing that the strongest evolution to $z\sim2$ is in the typical luminosity/break of the luminosity function, which evolves significantly. However, individual measurements for the UV luminosity function at $z<2$ are still  significantly affected by cosmic variance, small sample sizes and much more uncertain dust corrections, and thus the H$\alpha$ analysis provides a much stronger constraint on the evolution of star-forming galaxies up to $z\sim2.2$.

%
%
\begin{table*}
 \centering
  \caption{The luminosity function and star formation rate density evolution for $0.4<z<2.2$, as seen through a completely self-consistent analysis using HiZELS. The measurements are obtained at $z=0.4$, $0.84$, $1.47$ and $2.23$, correcting for 1 mag extinction at H$\alpha$. Columns present the redshift, break of the luminosity function, $L^*_{H\alpha}$, normalisation ($\phi^*_{H\alpha}$) and faint-end slope ($\alpha$) of the H$\alpha$ luminosity function. The two right columns present the star formation rate density at each redshift based on integrating the luminosity function down to $\approx3$\,M$_{\odot}$\,yr$^{-1}$ or 41.6 (in $\log$
    erg\,s$^{-1}$) and for a full integration. Star formation rate densities include a correction for AGN contamination of 10\% at $z=0.4$ and $z=0.84$ \citep[see][]{Garn2010a} and 15\% at both $z=1.47$ and $z=2.23$. Errors on the faint-end slope $\alpha$ are the 1\,$\sigma$ deviation from the best fit, when fitting the 3 parameters simultaneously. As $\alpha$ is very well-constrained at all redshifts, and shown not to evolve significantly, $L^*_{\rm H\alpha}$ and $\phi^*_{H\alpha}$ are obtained by fixing $\alpha=-1.6$, and the 1\,$\sigma$ errors on $L^*_{H\alpha}$ and $\phi^*_{\rm H\alpha}$ are derived from such fits (with fixed $\alpha$). }
  \begin{tabular}{@{}ccccccccc@{}}
  \hline
   \bf Epoch & $\bf L^*_{\rm H\alpha}$ & $\rm \bf \phi^*_{\rm \bf H\alpha}$ & $\bf \alpha_{\rm \bf H\alpha}$ & \bf log\,$\rho_{\bf L_{\bf H\alpha}}$\bf 41.6  & \bf log\,$\rho_{\bf L_{\bf H\alpha}}$  & $\bf \rho_{\rm \bf SFR H\alpha}$ \bf 41.6  & $\bf \rho_{\rm \bf SFR H\alpha}$ \bf All  \\
     ($z$)    & erg\,s$^{-1}$ & Mpc$^{-3}$ &    & erg\,s$^{-1}$\,Mpc$^{-3}$ & erg\,s$^{-1}$\,Mpc$^{-3}$ & M$_{\odot}$\,yr$^{-1}$ Mpc$^{-3}$   & M$_{\odot}$\,yr$^{-1}$ Mpc$^{-3}$  \\
 \hline
   \noalign{\smallskip}
$z=0.40\pm0.01$ & $41.95^{+0.47}_{-0.12}$ & $-3.12^{+0.10}_{-0.34}$ & $-1.75^{+0.12}_{-0.08}$ & $38.99^{+0.19}_{-0.22}$ & $39.55^{+0.22}_{-0.22}$ & $0.008^{+0.002}_{-0.002}$  & $0.03^{+0.01}_{-0.01}$ \\
$z=0.84\pm0.02$& $42.25^{+0.07}_{-0.05}$ & $-2.47^{+0.07}_{-0.08}$ & $-1.56^{+0.13}_{-0.14}$ & $39.75^{+0.12}_{-0.05}$ & $40.13^{+0.24}_{-0.21}$ & $0.040^{+0.007}_{-0.006}$  & $0.10^{+0.01}_{-0.02}$ \\
$z=1.47\pm0.02$ & $42.56^{+0.06}_{-0.05}$ & $-2.61^{+0.08}_{-0.09}$ & $-1.62^{+0.25}_{-0.29}$ & $40.03^{+0.08}_{-0.07}$ & $40.29^{+0.16}_{-0.14}$ & $0.07^{+0.01}_{-0.01}$  & $0.13^{+0.02}_{-0.02}$ \\
 $z=2.23\pm0.02$ & $42.87^{+0.08}_{-0.06}$ & $-2.78^{+0.08}_{-0.09}$ & $-1.59^{+0.12}_{-0.13}$ & $40.26^{+0.01}_{-0.02}$ & $40.44^{+0.03}_{-0.03}$ & $0.13^{+0.01}_{-0.01}$  & $0.21^{+0.02}_{-0.03}$ \\
 \hline
\end{tabular}
\label{lfs__}
\end{table*}

\section{The Star formation history of the Universe: the last 11 Gyrs with H$\alpha$}  \label{SFRD_z147}

Unveiling the star formation history of the Universe is fundamental to understand how, when and at what pace galaxies assembled their stellar masses. The best-fit Schechter function fit to the H$\alpha$ luminosity functions at $z=0.4$, $0.84$, $1.47$ and $2.23$ can be used to estimate the star formation rate density at the four epochs, corresponding to look-back times of 4.2, 7.0, 9.2 and 10.6 Gyrs. The standard calibration of Kennicutt (1998) is used to convert the extinction-corrected H$\alpha$ luminosity to a star formation rate:
\begin{equation}
{\rm SFR}({\rm M}_{\odot} {\rm yr^{-1}})= 7.9\times 10^{-42} ~{\rm L}_{\rm H\alpha} ~ ({\rm erg\,s}^{-1}),
\end{equation}
which assumes continuous star formation, Case B recombination at $T_e = 10^4$\,K and a Salpeter initial mass function ranging from 0.1--100\,M$_{\odot}$.

As detailed before, a constant 1 magnitude of extinction at H$\alpha$ is assumed for the analysis, which is likely to be a good approach for the entire integrated star formation.

\subsection{Removal of the AGN contribution}  \label{AGN}

Interpreting the integral of the H$\alpha$ luminosity function as a star formation rate density requires a good estimation of the possible contribution of AGN to that quantity. For the $z=0.84$ sample, \cite{Garn2010a} conducted a detailed search for potential AGN, finding a fraction of $8\pm3$\% within the H$\alpha$ population at $z=0.84$. Similar (i.e., $\sim10$\,\%) AGN contaminations at lower redshift have also been found by other studies, and therefore assuming a 10\% contribution from AGN up to $z\sim1$ is likely to be a good approximation. At higher redshifts, and particularly for the sample at $z=1.47$ and $z=2.23$, the AGN activity could in principle be different. By looking at a range of AGN indicators -- X-rays, radio and {\sc irac} colours (and emission lines ratios for sources with such information\footnote{Follow up spectroscopy of luminous H$\alpha$ sources unveil a significant fraction of narrow and broad-line AGN which becomes dominant at the highest luminosities (Sobral et al. in prep), but is consistent with an overall AGN contribution to the H$\alpha$ luminosity density of 15 per cent.}), it is found that $\sim15$\,\% of the sources are potentially AGN at $z=1.47$. Similar results are found at $z=2.23$. Therefore, when converting integrated luminosities to star formation rate densities at each epoch, it is assumed that AGNs contribute 10\% of that up to $z\sim1$ and 15\% above that redshift. While this correction may be uncertain, the actual correction will likely be within 5\% of what is assumed, and in order to guarantee the robustness of the measurements, the final measurements include the error introduced by the AGN correction -- this is done by adding 20\% of the AGN correction in quadrature to the other errors. The AGN contribution/contamination will be studied in detail in Sobral et al. (in prep.).

%
%
\begin{figure*}
\centering
\includegraphics[width=12.2cm]{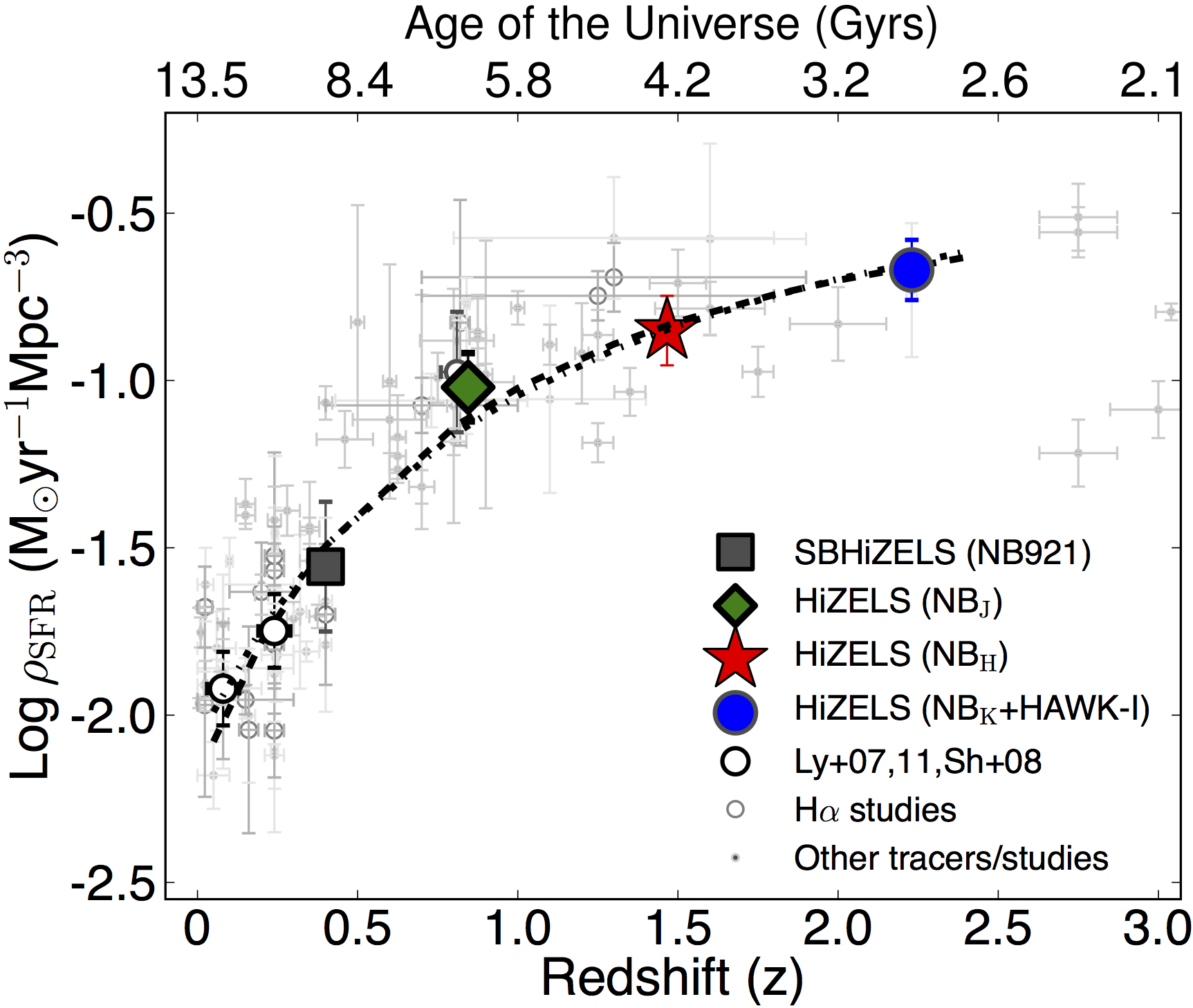}
\caption[SFRD]{The star formation rate density and its evolution with redshift up to $z\sim2.3$ using H$\alpha$ only, and compared to estimates at different redshifts from the literature \citep[e.g.][and references therein]{Hopkins2004,Shioya,CHU11}. This confirms a strong evolution in the star formation rate density over the last $\sim11$\,Gyrs. Overall, the simple parameterisations $\log\rho_{\rm SFR}=-0.14T-0.23$, with T in Gyrs (shown as a dashed line) or 2.1/($z$+1), shown as a dot-dashed line, provide good approximations to the star formation history of the Universe in the last 11\,Gyrs. This is in very good agreement with results from Karim et al. (2011), using radio stacking over a similar redshift range in the COSMOS field. \label{SFRD_HISTORY}}
\end{figure*}

%
%
\begin{figure*}
\centering
\includegraphics[width=12.5cm]{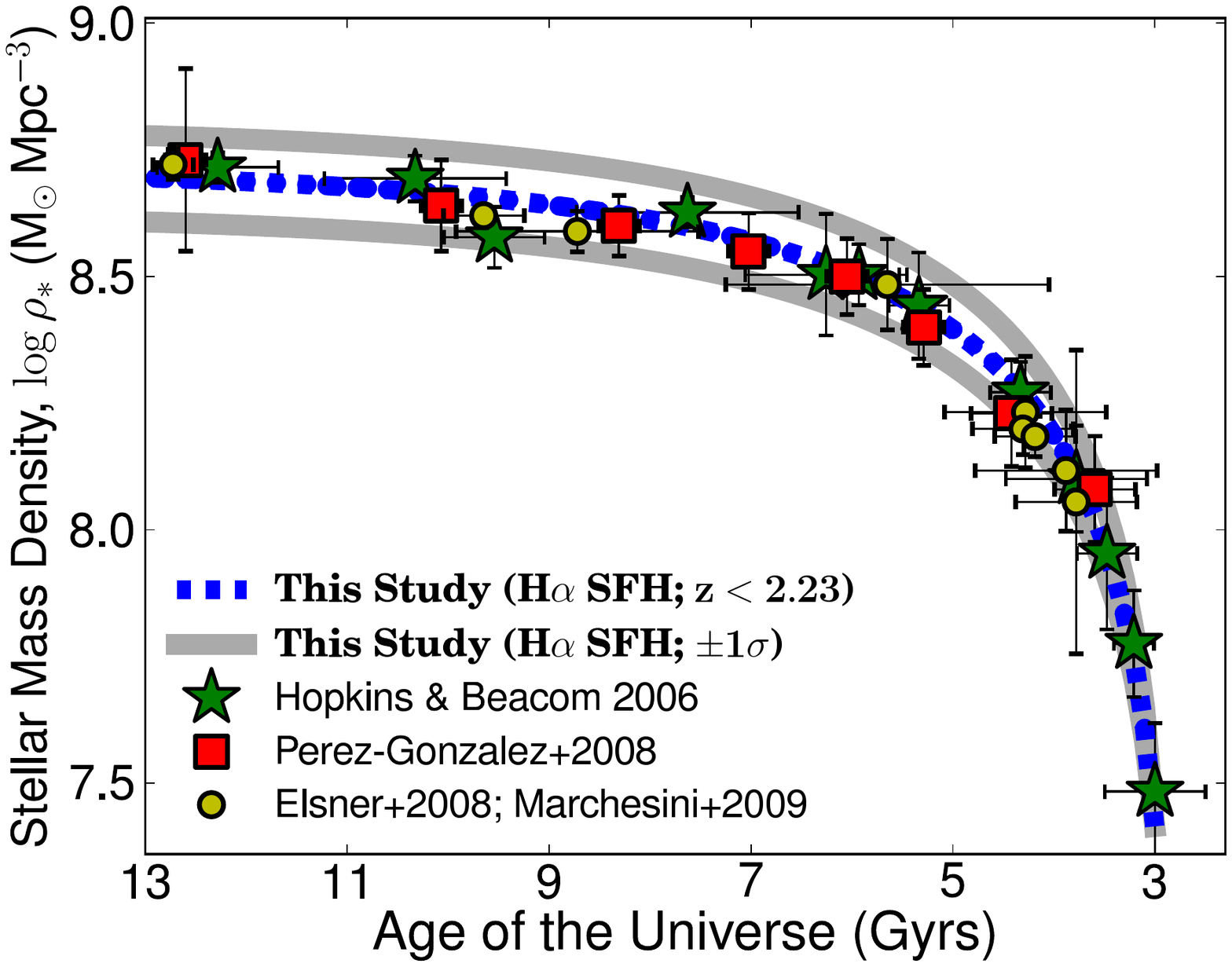}
\caption[SFRD]{The Stellar Mass assembly growth inferred from the completely self-consistent H$\alpha$ star formation history of the Universe over the last $\sim11$\,Gyrs, and the comparison with the observations of the stellar mass density over the same epochs. The results show a good agreement, suggesting that the H$\alpha$ analysis is indeed recovering essentially all the cosmic star formation happening since $z=2.23$.\label{Mass_Assembly}}
\end{figure*}

\subsection{The H$\alpha$ Star Formation History of the Universe}  \label{SFHISTORY}

The results are shown in Table 5, both down to the approximate common survey limits, and by fully integrating the luminosity function. Figure \ref{SFRD_HISTORY} also presents the results (fully integrating down the luminosity functions), and includes a comparison between the consistent view on the H$\alpha$ star formation history of the Universe derived in this paper with the various measurements from the literature \citep[][]{Hopkins2006,CHU11}, showing a good agreement.

The improvement when compared to other studies is driven by: i) the completely self-consistent determinations, ii) the significantly larger samples, and iii) the fact that the faint-end slope is accurately measured from $z\sim0$ to $z\sim2.23$ and luminosity functions determined down to a much lower common luminosity limit than ever done before. A comparison with all other previous measurements (which show a large scatter) reveals a good agreement with the H$\alpha$ measurements. However, the homogeneous H$\alpha$ analysis provides, for the first time, a much clearer and cleaner view of the evolution. 
The results presented in Figure \ref{SFRD_HISTORY} reveal the H$\alpha$ star formation history of the Universe for the last $\sim11$\,Gyrs. The evolution is particularly steep up to about $z\sim1$. While the evolution is then milder, $\rho_{\rm SFR}$ continues to rise, up to at least $z\sim2$.

Up to $z\sim1$, the H$\alpha$ star formation history is well fitted by $\log\rho_{\rm SFR}=4\times(z+1)-2.08$. However, such parameterisation is not a good fit for higher redshifts. It is possible to fit the entire H$\alpha$ star formation history since $z\sim2.2$, or for the last 11\,Gyrs by the simple parameterisation $\log\rho_{\rm SFR}=-0.14T-0.23$, with $T$ being the time since the Big Bang in Gyrs (see Figure 10). A power-law parameterisation as a function of redshift ($a\times(1+z)^{\beta})$ yields $\beta=-1.0$, and thus the H$\alpha$ star formation history can also be simply parameterised by $\log\rho_{\rm SFR}=\frac{-2.1}{(z+1)}$, clearly revealing that $\rho_{\rm SFR}$ has been declining for the last $\sim11$\,Gyrs. This parameterisation is also a very good fit for results from Karim et al. (2011), using radio stacking over a similar redshift range in the COSMOS field.

\subsection{The Stellar mass assembled in the last 11 Gyrs}  \label{MASS_ASSEMBLED}

The results presented in this paper can be used to provide an estimate of the stellar mass (density) which has been assembled by H$\alpha$ star-forming galaxies over the last 11\,Gyrs. This is done in a similar way to \cite{Hopkins2006} or \cite{Glazebrook04}, taking into account that a significant part of the mass of newborn stars at each redshift is recycled, and can be used in subsequent star formation episodes. The fraction of recycled mass depends on the IMF used. For a Salpeter IMF, which has been used for the H$\alpha$ calibration, the recycling fraction is 30\%. Note, however, that changing the IMF does not change the qualitative results presented in this paper, in particular the agreement between the predicted and the measured stellar mass density growth. Nevertheless, changing the IMF changes both the normalisation of the star formation history and the stellar mass density growth.

Here, the following approach is taken: the measured stellar mass density already in place at $z\sim2.2$ \citep[many determinations exist, e.g.][]{Hopkins2006, PerezGonz,Ilbert09} is assumed to be $\log_{10}\,M=7.45$\,M$_{\odot}$\,Mpc$^{-3}$. By using the measured H$\alpha$ star formation history derived in this paper ($\log\rho_{\rm SFR}=-0.14T-0.23$), a prediction of the evolution of the stellar mass density of the Universe is computed, using the recycling fraction of the Salpeter IMF (30\%).

The results are presented in Figure \ref{Mass_Assembly}, and compared with various measurements of the stellar mass density at different redshifts available from the literature \citep[][]{Hopkins2006,PerezGonz,Elsner08,Marchesini09}. All literature results  have been converted to a Salpeter IMF if derived with a different IMF -- including those with a modified Salpeter IMF (SalA; resulting in masses a factor of 0.77 lower than Salpeter; see e.g. Hopkins \& Beacom). 

The results reveal a very good agreement between the predictions based on the H$\alpha$ star formation history of the Universe presented in this paper since $z=2.23$ and the stellar mass density evolution of the Universe, measured directly by many authors. The results therefore indicate that at least since $z=2.23$ the H$\alpha$ star formation history of the Universe is a very good representation of the total star formation history of the Universe. It is possible to reconcile the observed evolution of the stellar mass density with that produced from the observed star formation history with very simple assumptions, without the need to modify the IMF or have it evolve as a function of time. The H$\alpha$ analysis reveals that star formation since $z=2.23$ is responsible for 95\% of the total stellar mass density observed today, with about half of that being assembled from $z\sim2.2$ to $z\sim1.2$, and the other half since $z\approx1.2$. Note that the same conclusion is reached if the stellar mass density at $z=0$ is adopted for the normalisation (instead of that at $z=2.23$), and the measured H$\alpha$ star formation history is used (with appropriate recycling factor) to evolve this stellar mass density back to earlier epochs. Moreover, if the star formation rate density continues to decline with time in the same way as in the last $\sim11$\,Gyrs, the stellar mass density growth will become increasingly slower, with the stellar mass density of the Universe reaching a maximum which is only 5\% higher than currently.


\section{Conclusions}  \label{conclusions}

This paper presents new results from a unique combination of wide and deep narrow-band H$\alpha$ surveys using UKIRT, Subaru and the VLT. It has resulted in robust and equally selected samples of several hundreds of H$\alpha$ emitters in narrow redshift slices, allowing to study and parameterise in a completely self-consistent way the evolution of the H$\alpha$ luminosity function over the last 11 Gyrs of the Universe. The main results are:

\begin{itemize}

\item We robustly select a total of 1742, 637, 515 and 807 H$\alpha$ emitters ($\Sigma$$>$3, EW$_{0\rm (H\alpha)}$$>$25\,\AA) across the COSMOS and the UDS fields at $z=0.40$, $0.84$, $1.47$ and $2.23$, respectively. These are by far the largest samples of homogeneously selected H$\alpha$ emitters, while the wide area and the coverage over two independent fields allows to greatly overcome cosmic variance and also assemble large samples of more luminous galaxies.

\item We find that the H$\alpha$ luminosity function evolves significantly from $z\sim0$ to $z\sim2.2$, with the bulk of the evolution being driven by the continuous rise in $L^*_{\rm H\alpha}$ by a factor of 10 from the local Universe to $z\sim2.2$, which is well described by $\log\,L^*_{\rm H\alpha}(z)=0.45z+41.87$

\item By obtaining very deep data over a wide range of epochs, it is found that the faint-end slope, $\alpha$ does not evolve with redshift up to $z\sim2.3$, and is set to $\alpha=-1.60\pm0.08$ for the last 11\,Gyrs ($0<z<2.2$), contrarily to previous claims (based on heterogeneous  samples) which argued for a steepening with redshift.

\item The evolution seen in the H$\alpha$ luminosity function is in good agreement with the evolution seen using inhomogeneous compilations of other tracers of star formation, such as FIR and UV, jointly pointing towards the bulk of the evolution in the last 11\,Gyrs being driven by a similar star-forming population across cosmic time, but with a strong luminosity increase from $z\sim0$ to $z\sim2.2$. 

\item This is the first time H$\alpha$ has been used to trace SF activity with a single homogeneous survey at $z=0.4-2.23$. The simple parametrisations $\log\rho_{\rm SFR}=-0.14T-0.23$ (with $T$ being the age of the Universe in Gyrs) or $\log\rho_{\rm SFR}=\frac{-2.1}{(z+1)}$ are good approximations for the last 11\,Gyrs, showing that $\rho_{\rm SFR}$ has been declining since $z\sim2.2$.
  
\item The results reveal that both the shape and normalisation of the H$\alpha$ star formation history are consistent with the measurements of the stellar mass density growth, confirming that the H$\alpha$ cosmic star formation history is tracing the bulk of the formation of stars in the Universe for $z<2.3$.

\item  The star formation activity over the last $\approx$\,11\,Gyrs is responsible for producing $\sim95$\% of the total stellar mass density observed locally today, with about half of that being assembled from $z\sim2.2$ to $z\sim1.2$, and the other half at $z<1.2$.

\end{itemize}

The results presented in this paper provide a self-consistent view that improves our understanding of the evolution of star-forming galaxies. Particularly, it shows that the evolution of the star-forming population in the last 11 Gyrs has been mostly driven by a change in the typical star formation rate of the population ($L^*_{\rm H\alpha}$), while the faint-end slope of the H$\alpha$ LF has remained constant ($\alpha=-1.6$), and the change in the normalisation has been much more moderate. The strong evolution in $L^*_{\rm H\alpha}$ (or SFR$^*$) may well be unveiling something very fundamental about the evolution of star-forming galaxies, as it seems to mark a transition between disk and mergers (e.g. S09) in the last 9-10\,Gyrs. Also, scaling SFRs by the SFR$^*$ at each epoch (or H$\alpha$ luminosities by $L^*_{\rm H\alpha}$ at each epoch) seems to recover relatively non-evolving relations between scaled SFRs/luminosities and e.g. dust extinction (S12), morphological class (S09), merger rates (Stott et al., 2012), or the typical dark matter halo in which the star-forming galaxies are likely to reside \citep{SOBRAL10A}. 

The results presented in this paper also complement the current view on the evolution of the stellar mass function over the last 11\,Gyrs \citep[e.g.][]{Ilbert09,Peng,Marchesini12}, which also reveal a non-evolving faint-end slope (of the stellar mass function) at least for $z<2$, but shallower, $\alpha=-1.3$. However, the typical mass of the stellar mass function, $M^*$, is found to be roughly constant in the last 11\,Gyrs, with the main change being $\phi^*$, which continuously increases in the last $\sim11$\,Gyrs. Combining the results of the evolution of the H$\alpha$ luminosity function with those of the evolution of the stellar mass function point towards the existence of a star-forming population which is mostly evolving by an overall decrease in their SFRs/luminosity, while the overall population of galaxies evolves by a change in number density, but with a rather non-evolving typical mass ($M^*$), a rather simple evolution scenario which is consistent with that proposed by \cite{Peng}.

\section*{Acknowledgments}

The authors would like to thank the reviewer, James Colbert, for many comments and suggestions which improved the paper significantly. DS is supported by a NOVA fellowship. IRS acknowledges a Leverhulme Senior Fellowship. PNB acknowledges support from the Leverhulme Trust. YM, JPS and IRS thank the U.K. Science and Technology Facility Council (STFC). JEG is supported by a Banting Fellowship, administered by the Natural Science and Engineering Research Council of Canada.  We would like to thank Richard Ellis, Simon Lilly, Peter Capak, Adam Muzzin, Taddy Kodama, Masao Hayashi, Andy Lawrence, Joop Schaye, Marijn Franx, Huub R\"ottgering, Rychard Bouwens and Renske Smit for many interesting and helpful discussions. We would also like to thank Chun Ly for helpful comments. The authors wish to thank all the JAC staff for their help conducting the observations at the UKIRT telescope, and their continuous support and we are also extremely grateful to all the Subaru staff. We also acknowledge ESO and Subaru for service observations. Finally, the authors fully acknowledge the tremendous work that has been done by both COSMOS and UKIDSS UDS/SXDF teams in assembling such large, state-of-the-art multi-wavelength data-sets over such wide areas, as those have been crucial for the results presented in this paper.

\bibliographystyle{mn2e.bst}
\bibliography{bibliography.bib}

\appendix

\section{Catalogues of candidate HiZELS narrow-band emitters}  \label{Cats}

The catalogues of potential narrow-band emitters over the COSMOS and UDS fields are presented in Tables A.1 (NB921), A.2 (NB$_{\rm J}$), A.3 (NB$_{\rm H}$) and A.4 (NB$_{\rm K}$). It contains IDs (including field and observing band), Right Ascension (RA), Declination (Dec), narrow-band magnitude (NB), broad-band magnitude (BB), the significance of the narrow-band excess ($\Sigma$), Estimated Flux ($\log10$), estimated observed EW, and a flag for those that are classified as H$\alpha$. Note that only the online version contains the full catalogue -- here only five entries of the table are shown as examples of the entire catalogues.

\begin{table*}
 \centering
  \caption{A catalogue of all $\Sigma>3$ NB921 narrow-band sources selected in the UDS and COSMOS fields from HiZELS. NB magnitudes are NB921 (AB); BB are $z'$ magnitudes (AB). Note that on the printed version this only contains 5 entries, in order to provide some examples contained in the full catalogue. The full catalogue is available on-line.}
  \begin{tabular}{@{}ccccccccccc@{}}
  \hline
   ID & R.A. & Dec. & NB & BB & $\Sigma$  & log Flux  & EW$_{\rm obs}$ & Class. as H$\alpha$  \\
         & {(J2000)} &(J2000) & (AB) &(AB) &  & erg\,s$^{-1}$ & \AA\\
 \hline
   \noalign{\smallskip}
HiZELS-COSMOS-NB921-S12-14 & 09\,58\,13.57 & $+$02\,17\,15.6 & 23.86$\pm0.11$ & 24.85$\pm0.08$ & 13.6 & $-16.255$ & 418.4 & No \\
HiZELS-COSMOS-NB921-S12-105 & 09\,59\,57.96 & $+$02\,17\,41.5 & 23.14$\pm0.03$ & 23.39$\pm0.02$ & 9.9 & $-16.132$ & 119.8 & No \\
HiZELS-COSMOS-NB921-S12-136 & 09\,58\,57.55 & $+$02\,17\,45.0 & 21.64$\pm0.01$ &  22.11$\pm0.01$ & 56.5 & $-15.636$ & 105.9 & Yes \\
HiZELS-COSMOS-NB921-S12-4064 & 09\,59\,20.49 & $+$02\,20\,54.5 & 22.06$\pm0.01$ & 22.21$\pm0.01$ & 14.3 & $-16.084$ & 39.13 & No \\
HiZELS-UDS-NB921-S12-220582  & 02\,16\,19.37 & $-$05\,13\,54.2 & 22.37$\pm0.01$ & 22.65$\pm0.01$ & 55.6 & $-16.287$ & 38.22 & Yes \\
 \hline
\end{tabular}
\label{NB921_CAT}
\end{table*}

\begin{table*}
 \centering
  \caption{A catalogue of all $\Sigma>3$ NB$_{\rm J}$ narrow-band sources selected in the UDS and COSMOS fields from HiZELS. NB magnitudes are NB$_{\rm J}$ (Vega); BB are $J$ magnitudes (Vega). Note that on the printed version this only contains 5 example entries, in order to provide some examples contained in the full catalogue. The full catalogue is available on-line.}
  \begin{tabular}{@{}ccccccccccc@{}}
  \hline
   ID & R.A. & Dec. & NB & BB & $\Sigma$  & log Flux  & EW$_{\rm obs}$ & Class. as H$\alpha$  \\
         & {(J2000)} &(J2000) & (Vega) &(Vega) &  & erg\,s$^{-1}$ & \AA\\
 \hline
   \noalign{\smallskip}
HiZELS-COSMOS-NBJ-S12-206 & 09\,58\,42.11 & $+$01\,58\,40.5 & 21.58$\pm0.20$ & 22.83$\pm0.23$ & 3.5 & $-16.11$ & 463.5 & No \\
HiZELS-COSMOS-NBJ-S12-292 & 09\,58\,42.65 & $+$02\,27\,24.2 & 20.51$\pm0.08$ & 20.85$\pm0.05$ & 3.7 & $-16.08$ & 65.4 & Yes \\
HiZELS-COSMOS-NBJ-S12-293 & 09\,58\,42.63 & $+$02\,19\,53.2 & 20.84$\pm0.10$ & 21.46$\pm0.07$ & 4.4 & $-16.01$ & 139.3 & Yes \\
HiZELS-UDS-NBJ-S12-5 & 02\,16\,16.81 & $-$05\,09\,09.2 & 20.51$\pm0.08$ & 20.85$\pm0.05$ & 5.4 & $-15.89$ &194.6 & Yes \\
HiZELS-UDS-NBJ-S12-26 & 02\,16\,17.15 & $-$05\,07\,42.4 & 20.84$\pm0.10$ & 21.46$\pm0.07$ & 4.6 & $-15.96$ & 79.8 & Yes \\
 \hline
\end{tabular}
\label{NBJ_CAT}
\end{table*}

\begin{table*}
 \centering
  \caption{A catalogue of all $\Sigma>3$ NB$_{\rm H}$ narrow-band sources selected in the UDS and COSMOS fields from HiZELS. NB magnitudes are NB$_{\rm H}$ (Vega); BB are $H$ magnitudes (Vega). Note that on the printed version this only contains 5 example entries, in order to provide some examples contained in the full catalogue. The full catalogue is available on-line.}
  \begin{tabular}{@{}ccccccccccc@{}}
  \hline
   ID & R.A. & Dec. & NB & BB & $\Sigma$  & log Flux  & EW$_{\rm obs}$ & Class. as H$\alpha$  \\
         & {(J2000)} &(J2000) & (Vega) &(Vega) &  & erg\,s$^{-1}$ & \AA\\
 \hline
   \noalign{\smallskip}
HiZELS-COSMOS-NBH-S12-206 & 09\,57\,47.31 & $+$02\,02\,08.1 & 21.58$\pm0.14$ & 20.99$\pm0.05$ & 3.1 & $-15.98$ & 251.4 & No \\
HiZELS-COSMOS-NBH-S12-292 & 09\,57\,47.33 & $+$01\,51\,53.1 & 20.51$\pm0.02$ & 18.33$\pm0.01$ & 12.1 & $-15.40$ & 77.3 & No \\
HiZELS-COSMOS-NBH-S12-293 & 09\,57\,47.65 & $+$02\,16\,19.8 & 20.84$\pm0.08$ & 19.80$\pm0.02$ & 3.3 & $-15.87$ & 101.8 & No \\
HiZELS-UDS-NBH-S12-4012 & 02\,16\,44.17 & $-$04\,44\,53.0 & 20.04$\pm0.18$ & 20.93$\pm0.04$ & 3.0 & $-15.87$ & 308 & Yes \\
HiZELS-UDS-NBH-S12-4136 & 02\,16\,45.35 & $-$04\,07\,26.8 & 19.15$\pm0.08$ & 19.76$\pm0.02$ & 5.2 & $-15.64$ & 171 & Yes \\
 \hline
\end{tabular}
\label{NBH_CAT}
\end{table*}

\begin{table*}
 \centering
  \caption{A catalogue of all $\Sigma>3$ NB$_{\rm K}$ narrow-band sources selected in the UDS and COSMOS fields from HiZELS. NB magnitudes are NB$_{\rm K}$ (Vega); BB are $K$ magnitudes (Vega). Note that on the printed version this only contains 5 example entries, in order to provide some examples contained in the full catalogue. The full catalogue is available on-line.}
  \begin{tabular}{@{}ccccccccccc@{}}
  \hline
   ID & R.A. & Dec. & NB & BB & $\Sigma$  & log Flux  & EW$_{\rm obs}$ & Class. as H$\alpha$  \\
         & {(J2000)} &(J2000) & (Vega) &(Vega) &  & erg\,s$^{-1}$ & \AA\\
 \hline
   \noalign{\smallskip}
HiZELS-COSMOS-NBK-S12-2227 & 09\,58\,14.05 & $+$02\,34\,05.1 & 20.26$\pm0.23$ & 21.68$\pm0.16$ & 3.1 & $-16.31$ & 728 & Yes \\
HiZELS-COSMOS-NBK-S12-45390 & 10\,00\,59.58 & $+$02\,44\,35.9 & 19.24$\pm0.09$ & 19.92$\pm0.04$ & 5.4 & $-16.09$ & 208 & No \\
HiZELS-COSMOS-NBKS12-44193 & 10\,00\,55.39 & $+$01\,59\,55.1 & 19.02$\pm0.06$ & 19.68$\pm0.02$ & 7.6 & $-16.01$ & 198 & Yes \\
HiZELS-UDS-NBK-S12-15961 & 02\,18\,09.03 & $-$04\,47\,49.9 & 17.47$\pm0.02$ & 18.27$\pm0.01$ & 30.6 & $-15.33$ & 262 & Yes \\
HiZELS-UDS-NBK-S12-22618 & 02\,18\,53.02 & $-$05\,01\,07.7 & 19.36$\pm0.10$ & 20.02$\pm0.02$ & 4.5 & $-16.16$ & 196 & No \\
 \hline
\end{tabular}
\label{NBK_CAT}
\end{table*}

\bsp

\label{lastpage}

\end{document}